\documentclass[slac_one]{revtex4}
\usepackage{graphicx}
\usepackage{fancyhdr}
\newcommand{\Eslash}{\mbox{$E \kern-0.6em\slash$                }}
\newcommand{\etmiss}{\mbox{$\Eslash_{T}\!$                        }}

\def\MET{{\mbox{$E\kern-0.57em\raise0.19ex\hbox{/}_{T}$}}}
\def\met{{\mbox{$E\kern-0.57em\raise0.19ex\hbox{/}_{T}$}}}
\def\DZ{D\O\ }
\def\DZero{D\O\ }

\def\ifb{fb$^{-1}$}
\def\pp{$p\bar{p}$}

\def\lmet{$WH\rightarrow \ell\kern-0.45em\raise0.19ex\hbox{/} \nu b\bar{b}$}

\def\pzh{$p\bar{p}\rightarrow ZH \rightarrow  \nu\bar{\nu} b\bar{b} / \ell^+ \ell^- b\bar{b}$}

\def\pwww{$p\bar{p}\rightarrow WH \rightarrow WW^{+} W^{-}$}
\def\pwwww{$p\bar{p}\rightarrow WH \rightarrow WW W$}

\def\phww{$p\bar{p}\rightarrow H \rightarrow W^{+} W^{-}$}
\def\phwww{$p\bar{p}\rightarrow H \rightarrow WW$}
\def\hww{$H\rightarrow W^+ W^-$}
\def\hwww{$H\rightarrow WW$}
\def\hbb{$H\rightarrow b\bar{b}$}

\def \ttbar {\ensuremath{ \rm t \bar{t}             }}
\def\lmet{$WH\rightarrow \ell\kern-0.45em\raise0.19ex\hbox{/} \nu b\bar{b}$}

\def\ZHll{$ZH\rightarrow \ell^+ \ell^- b\bar{b}$}

\def\pwh{$p\bar{p}\rightarrow WH \rightarrow \ell \nu b\bar{b}$}
\def\pzh{$p\bar{p}\rightarrow ZH \rightarrow  \nu\bar{\nu} b\bar{b} / \ell^+\ell^- b\bar{b}$}

\def\pwww{$p\bar{p}\rightarrow WH \rightarrow WW^{+} W^{-}$}

\def\phww{$p\bar{p}\rightarrow H \rightarrow W^{+} W^{-}$}
\def\hww{$H\rightarrow W^+ W^-$}
\def\hbb{$H\rightarrow b\bar{b}$}

\begin{document}

\title{{\small{Hadron Collider Physics Symposium (HCP2008),
Galena, Illinois, USA}}\\ 
\vspace{12pt}
Searches and Prospects for Standard Model Higgs boson at the Tevatron} 

%

\author{Gregorio Bernardi}
\affiliation{LPNHE-Paris, Universities of Paris 6 and 7, France}
\author{on behalf of the CDF and D\O\ Collaborations}
\affiliation{ }

\begin{abstract}
We report on  results obtained at the Tevatron by the CDF and D\O\
collaborations up to June 2008 (2.4 fb$^{-1}$), 
on searches for standard model (SM) Higgs bosons having
a high mass (135-200 GeV). High mass  Higgs bosons decay dominantly in  $WW^{\star}$
and the presented searches are performed in the leptonic decay modes of the $W's$. Both
direct production ($p\bar{p} \rightarrow H$) and associated production 
($p\bar{p} \rightarrow WH$) are studied and eventually combined with all channels
available at the Tevatron.
Prospects for SM Higgs searches with the full projected Tevatron statistics are also given.
\end{abstract}

\maketitle

\thispagestyle{fancy}


\section{INTRODUCTION} 
The Tevatron experiments are searching for the standard model (SM) Higgs
boson which is expected to be the footprint of
the spontaneous breaking of the  electroweak symmetry. This mechanism
provides
an explanation for the masses of the elementary particles, which are massless
in the unbroken gauge theory, so experimental confirmation is eagerly awaited
for. Direct searches at LEP have constrained its mass to be above 
114.4 GeV~\cite{lepdirect}
while indirect searches contrain it to be below 190 GeV, when taking
into account the direct bound~\cite{lepindirect}.
We report here  on the searches at high mass (defined
as the searches in which the Higgs boson decays in a pair of $W's$)
while the low mass analyses are described in another contribution to this 
conference~\cite{hughes}. 
For Higgs boson searches, at ``high'' mass (135-200 GeV), 
the most sensitive  production channel at the Tevatron 
(center-of-mass energy of $\sqrt{s}=1.96\,\mbox{TeV}$)
is the direct production ($p \bar{p} \rightarrow H$),
but all possible channels are studied to gain sensitivity through their combination,
in particular associated Higgs-electroweak boson production ($p \bar{p} \rightarrow WH,ZH$).

In this report, we first describe briefly the $WH$ analyses, which have lower sensitivity at high mass,
then report in more details on the \hwww\ analyses. We then present the combination of these
results with those obtained on searches for low mass Higgs
and conclude with the prospects on SM Higgs boson search at the Tevatron.

\section{SEARCH FOR \ \pwwww }

The search for \pwwww\ relies on the search for two leptons  having the same electric charge
(in this report, ``lepton'' refers to  electron, $e$, or  muon, $\mu$). No explicit
requirements are put on the decay of the third $W$ produced, and the like-sign condition
ensures orthogonality with the \phwww\ searches, described below.

At CDF, a new analysis with 1.9 fb$^{-1}$ has recently been presented~\cite{cdf-www}.
For the event selection the requirements are:
at least one electron with $E_T >$ 20 GeV and $p_T >$ 10 GeV or one 
muon with $p_T >$ 20 GeV, and at least one other electron with $E_T >$ 6 GeV 
and $p_T >$ 6 GeV or one muon with $p_T >$ 6 GeV. Other lepton selection 
requirements include isolation, track quality, and the consistency of 
detector response with expectations for electrons or muons. Photon 
conversion resulting in electron candidates are rejected by identifying an 
oppositely charged track satisfying a conversion configuration. For the 
events containing exactly two leptons passing the lepton selection, 
the two leptons must originate from the same event vertex, satisfy
a di-lepton mass selection of ($M_{ll} >$ 12 GeV) and a $Z$-event 
veto. 
%
%

The remaining background can be classified in two types, "physics"
and "instrumental".
Backgrounds containing prompt real leptons (physics backgrounds) are 
estimated using simulated samples. Such backgrounds can be further classified 
into reducible (Drell-Yan, W+heavy flavor hadrons, $t \bar{t}$, and $WW$), 
and irreducible ($WZ, ZZ$).
Instrumental backgrounds include
  residual photon-conversion events which are one of the dominant 
backgrounds for such like-sign di-lepton analysis. Such background
is determined from the data and the simulation~\cite{cdf-www}.
  Other types of fake lepton events are also an important source of background,
 for instance the overlap of a charged and a
neutral pion, or the misidentification of a charged pion into a muon.
Their estimation is described in detail in~\cite{cdf-www}.
For the final selection the like-sign di-lepton events are examined
on the two dimensional plane of the 2$^{nd}$ lepton $p_T$ (pT2)  vs the 
di-lepton system $p_T$ (pT12). 
In the most sensitive region,
pT2 $>$ 20 GeV and pT12 $>$ 15 GeV,  3 events are observed, while
3.2 background events and 0.2 Higgs events are expected,
if $m_H=160$ GeV.
No significant 
discrepancies are found between the data and the background expectations 
and so four regions are selected in this plane from which the limits
are extracted and then combined. 

\begin{figure*}[htbp]
\centering
\includegraphics[width=90mm,height=70mm]{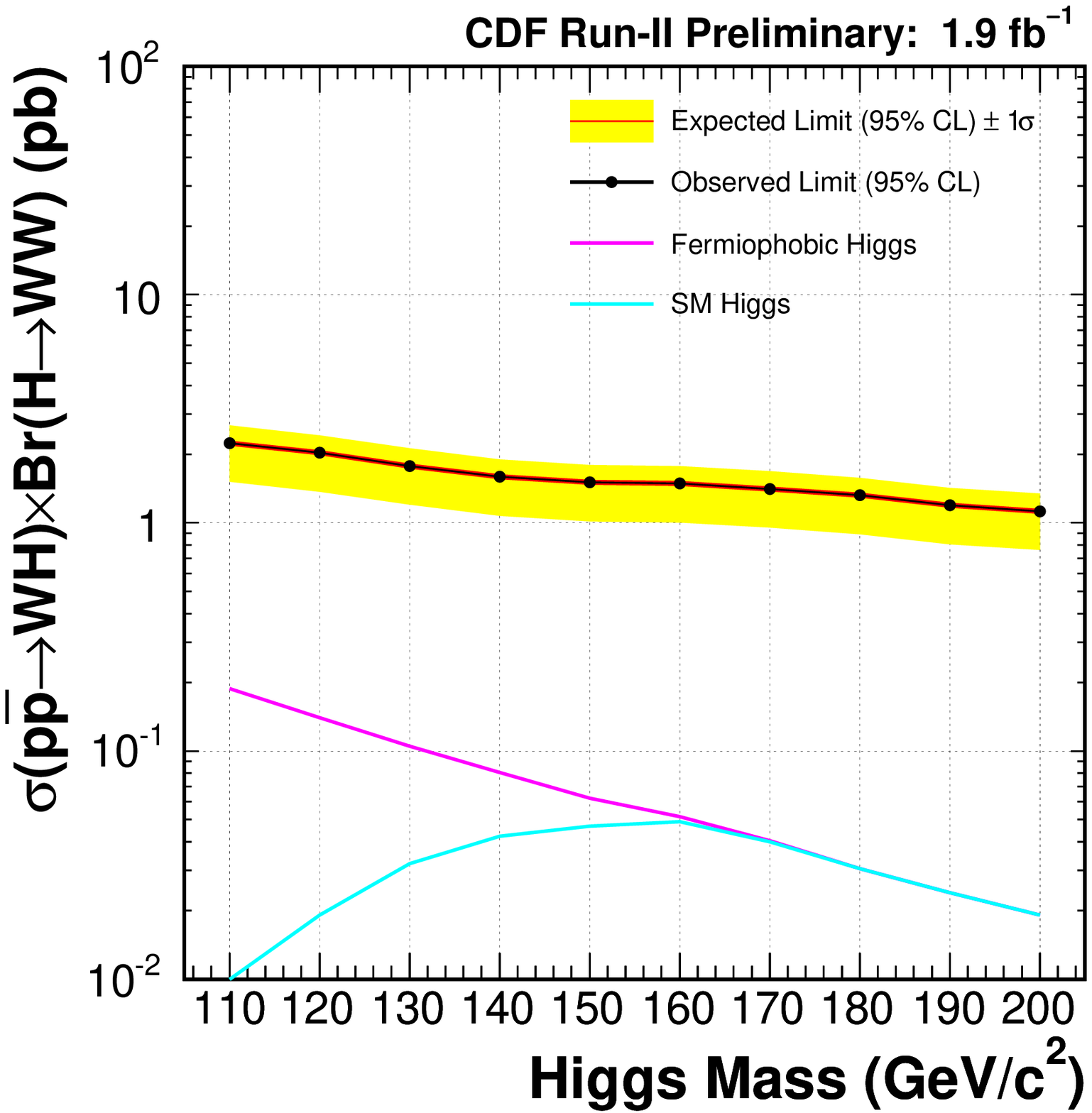}
\includegraphics[width=85mm]{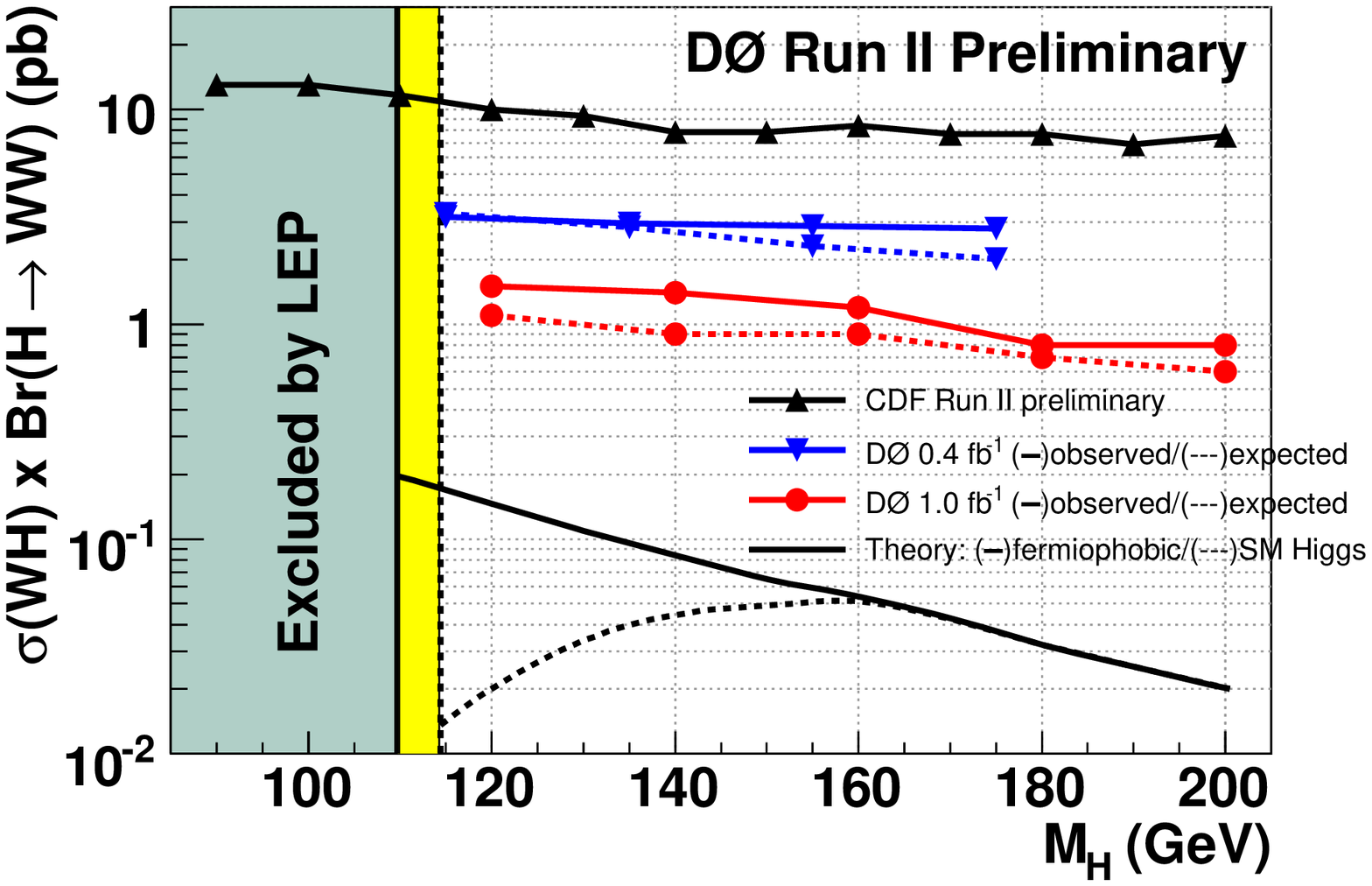}
\caption{95\% C.L. limits on $WH \rightarrow WWW$ production cross section, as 
obtained by CDF and D\O }
 \label{www-limits}
\begin{picture}(0,0)(0,0)
  \put (-205,175){\scriptsize {\bf (a)}}
  \put ( 48,175){\scriptsize {\bf (b)}}
\end{picture} 
\end{figure*}

The D\O\ collaboration did not update its 2007  result based on 1.1 fb$^{-1}$. 
The selection was similar to the one described above for the recent CDF result, and 
the instrumental background is also determined from the data but with a different 
technique~\cite{d0-www}.
An important difference with the CDF analysis comes from the limit determination which
in the case of D\O\ was set using the results
of a topological likelihood discriminant~\cite{d0-www-pub} resulting in limits of comparable
sensitivity, as shown in
figures~\ref{www-limits}. These limits are still a factor of about 20 above the standard model
prediction for $m_H=160$ GeV, but are more sensitive at low 
mass if one assumes fermiophobic models.

\section{SEARCH FOR \ \phwww }

The most sensitive channel at high mass is by far \phwww , where production
can be via gluon-gluon or vector-boson fusion. Both CDF and D\O\ 
have updated their previous results, and are using 2.4 and 2.3 fb$^{-1}$ 
respectively. 
We first describe the individual analysis, then their combination.

\subsection{\DZ Results on \ \phwww}
The $H \rightarrow W W$ 
 candidates are selected by triggering on single or
di-lepton events with approximately 100\% efficiency for the events
of the final selection~\cite{d0-ww}.
In the offline selection,
electrons must be reconstructed within a detector pseudorapidity
$|\eta| < 3.0$.  Muons are required to be in 
the fiducial coverage of the muon system $|\eta| < 2.0$.
The two leptons must originate from the same primary
vertex and are required to be of opposite charge. They must have
$p_T^{e} >$ 20~GeV for the leading electron and $p_T^{e} >$ 15~GeV
for the trailing one in the $ee$ channel, $p_T^{e} >$ 15~GeV for the
electron and $p_T^{\mu} >$ 10~GeV for the muon in the $e\mu$ final
state and $p_T^{\mu} >$ 10~GeV for the leading and trailing muons in
the $\mu\mu$ final state. The di-lepton invariant mass
must also  be greater than 15 GeV. 

At this  ``pre-selection'' stage, 
the background is dominated by $Z/\gamma^*$
production which is suppressed by requiring missing transverse
energy and scaled missing transverse energy~\cite{d0-ww}:
 $Z/\gamma^*$, di-boson 
 and multi-jet events are also rejected  with a cut on the
opening azimuthal angle $\Delta \varphi_{\ell\ell}$, since most of the background
decays are back-to-back which is not the case for Higgs boson decays
because of the spin correlations induced by its scalar nature.
$t\overline{t}$ events are further rejected by a cut on $H_T$, the
scalar sum of the $p_T$ of good jets in the event.

The signal and SM background processes are simulated with
\textsc{Pythia}~\cite{pythia} (except for $W$/$Z$+jets in the
$\mu\mu$ channel, where \textsc{ALPGEN}~\cite{alpgen} is used) 
using the CTEQ6.1M~\cite{cteq6} parton distribution
functions.
The $Z/\gamma^* \rightarrow
\ell\ell$ cross section is calculated 
at NNLO~\cite{hamberg}  with CTEQ6.1M PDFs.
The NLO $WW$, $WZ$ and $ZZ$  production cross
section values are taken from~\cite{ellis}. 
The background due to multijet production, when
jets are misidentified as leptons, is determined from the data.
See~\cite{d0-ww} for more details and for the simulation of the  other backgrounds.

Additional Higgs mass and final-state dependent selections are
optimized to further suppress contributions from $Z/\gamma^*$,
di-boson ($WW, WZ, ZZ$), $W(\rightarrow \ell\nu)+jets$, and multijet
backgrounds. Table~\ref{yields} shows the
number of expected and observed events after pre-selection and final
selections (i.e. NN input stage), for all three channels.

\begin{table}[htb]
\begin{center}
\begin{tabular}{|c|cc|cc|cc|}
\hline 
                & $ee$ pre-selection & $ee$ final & $e\mu$ pre-selection & $e\mu$ final & $\mu\mu$ pre-selection & $\mu\mu$ final \\
\hline Signal
($m_H=160$ GeV) & $1.34\pm0.03$ & $0.82\pm0.02$ & $3.58\pm0.05$ & $1.76\pm0.04$ & $3.52\pm0.08$   & $2.08 \pm 0.03$ \\
\hline
Total Background& $49306\pm306$ & $10.7\pm1.7\pm1.7$ & $1436\pm138$ & $21.1\pm1.3\pm3.4$ & $110681\pm231$ & $831.8\pm29\pm34$ \\
\hline
Data            & $50593$ & $10 $ & $1424$ & $18$ & $109918$ & $839$ \\
\hline
\end{tabular}
\caption{\label{yields} Expected and observed number of events in
each channel after pre-selection and final selections (NN input
stage). Statistical and systematic uncertainties in the expected yields are shown
for the final backgrounds.} 
\end{center}
\end{table}

\begin{figure*}[htbp]
\centering
\includegraphics[width=65mm]{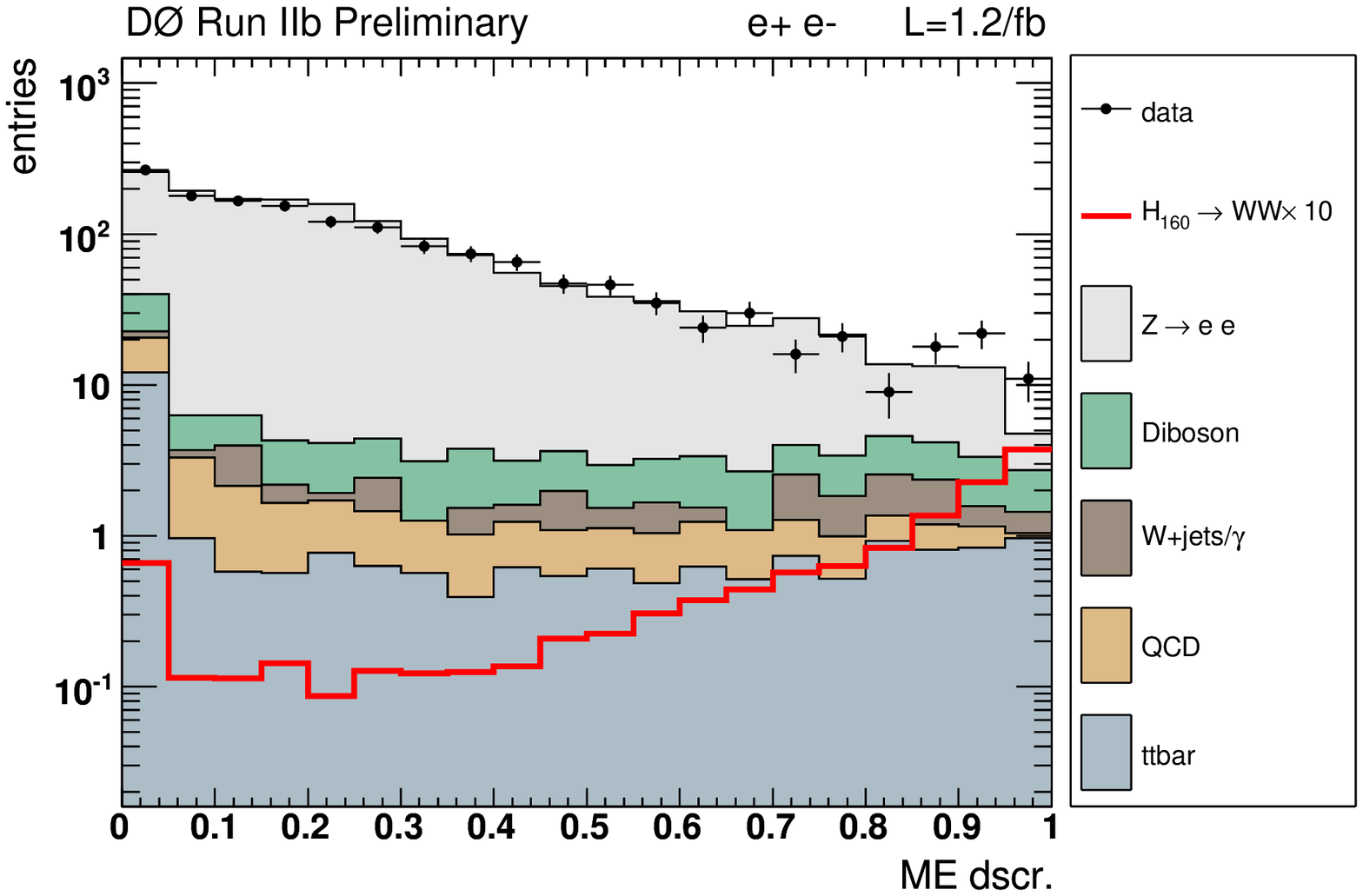}
\includegraphics[width=65mm]{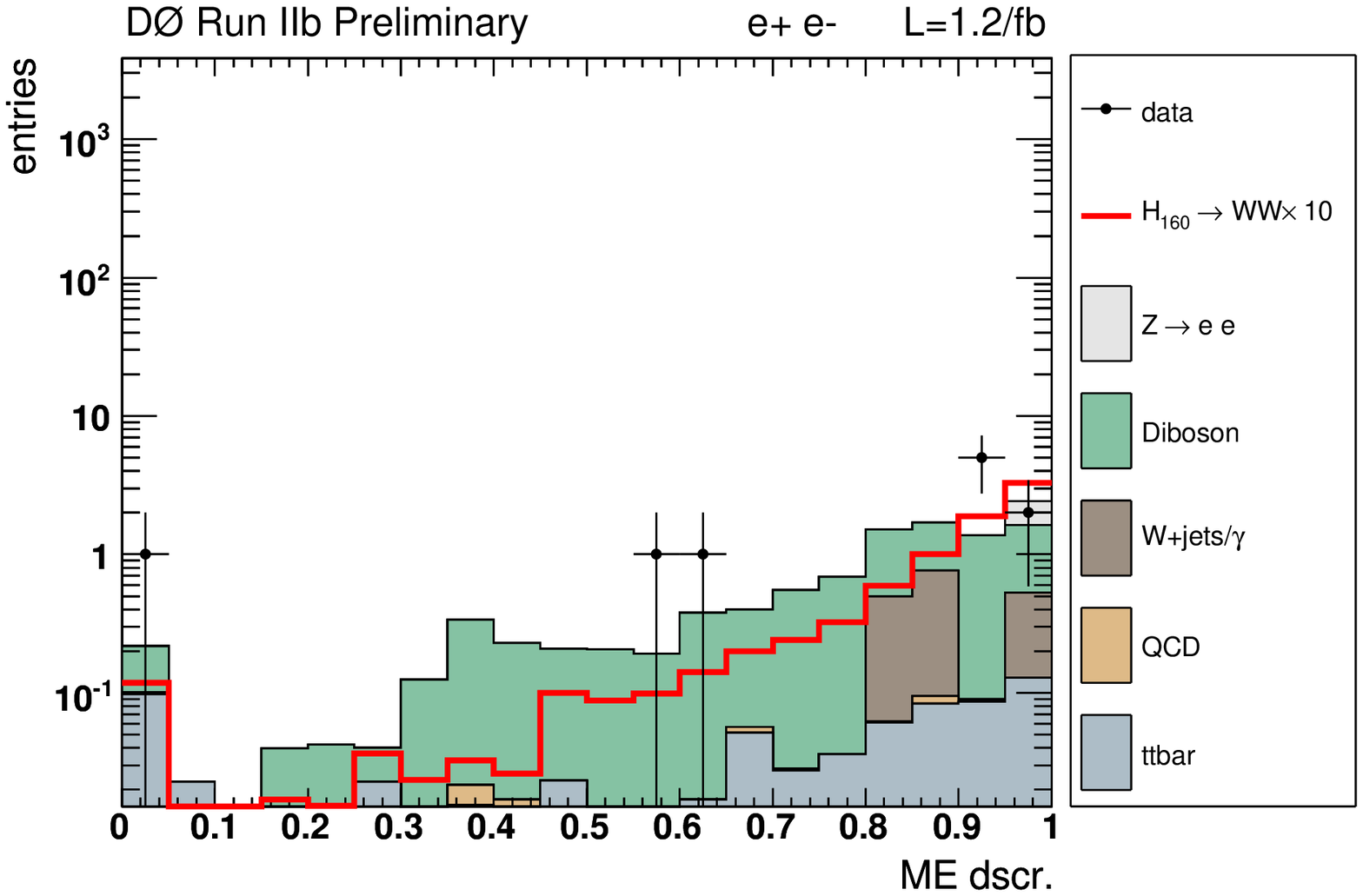}
\includegraphics[width=65mm]{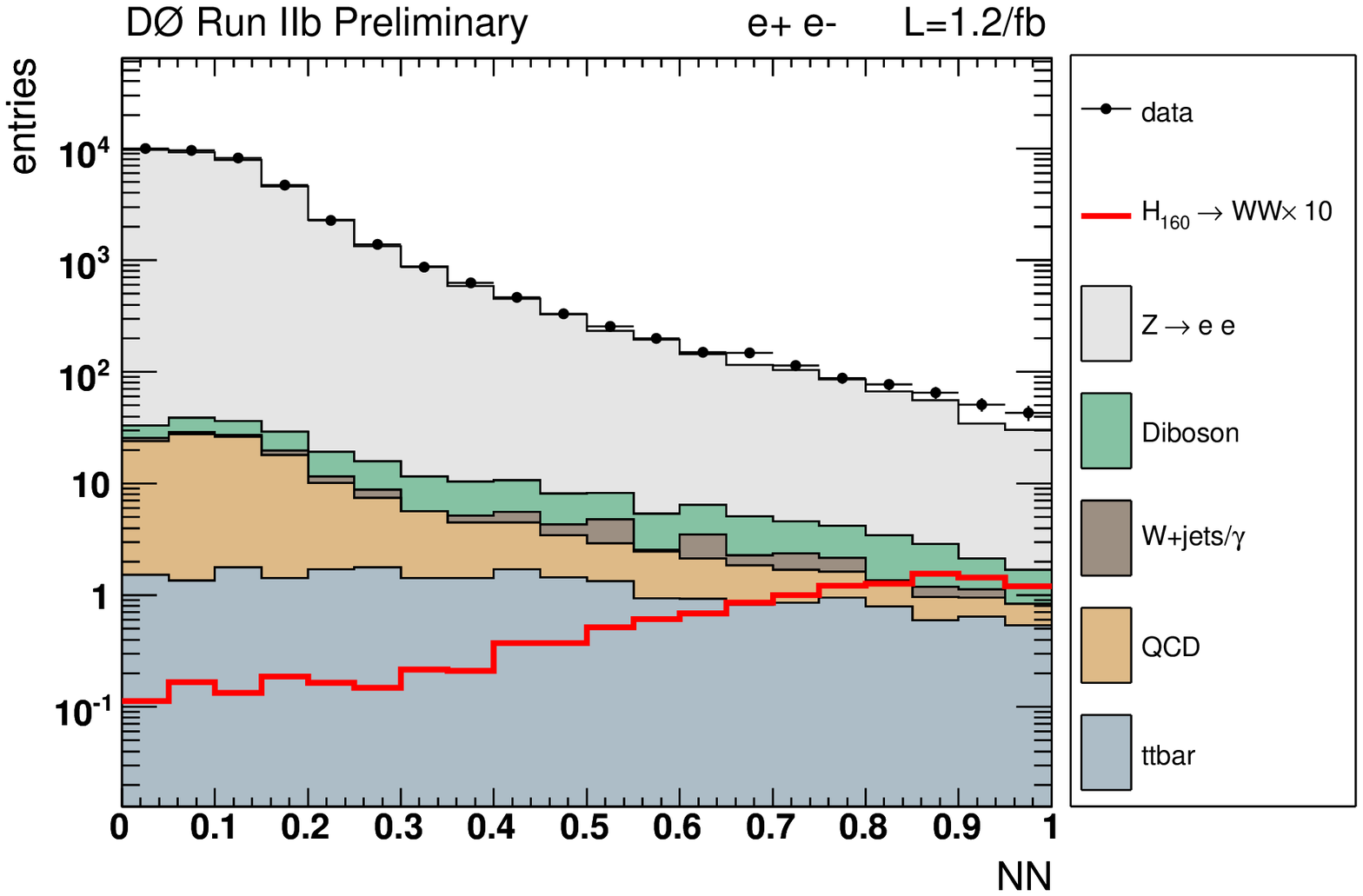}
\includegraphics[width=65mm]{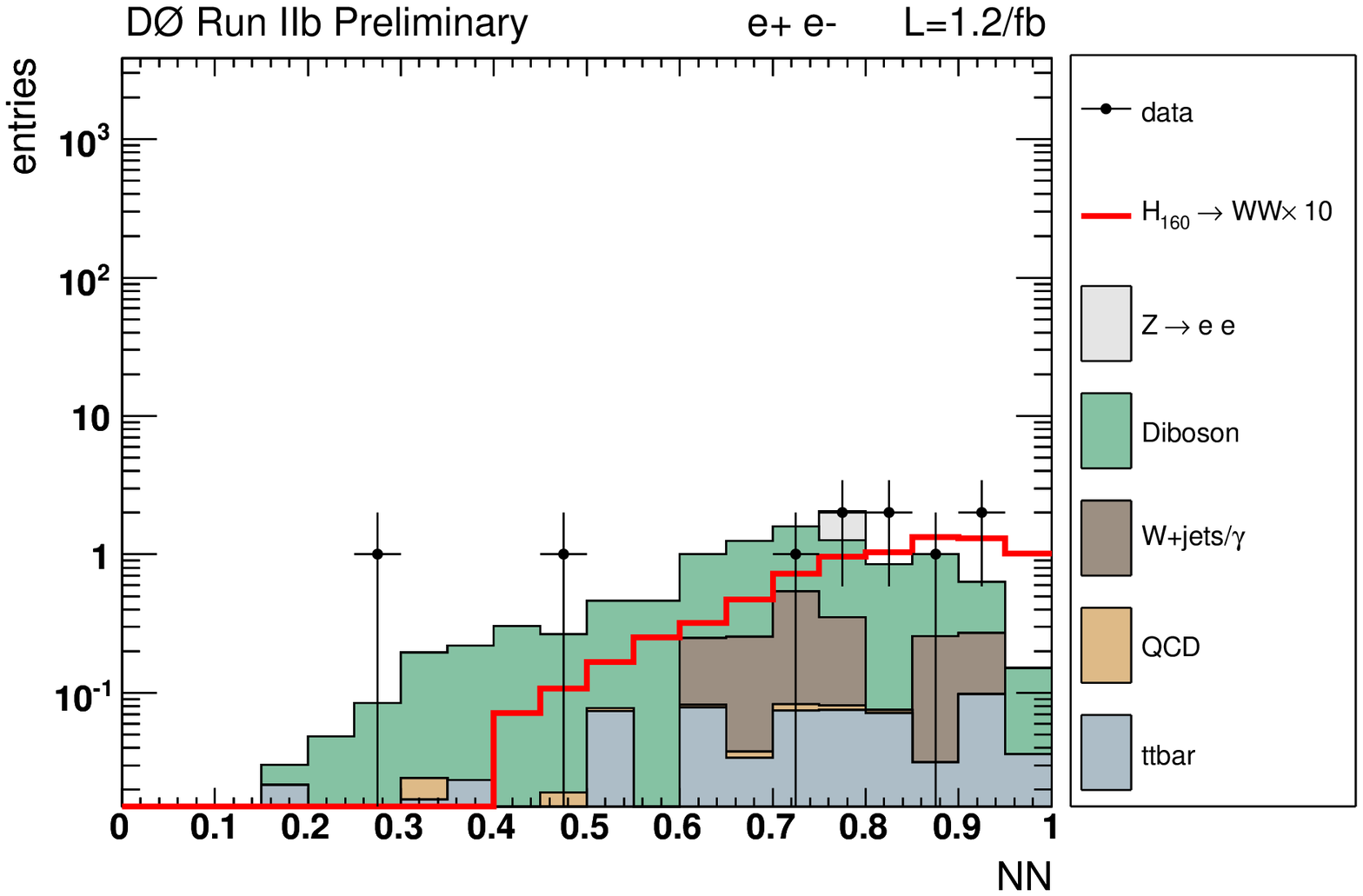}
\caption{Distribution of the ME and NN output in the D\O\ $ee$ analysis at the preselection (a and c) and 
at final selection level (b and d), in the Run IIb (1.2 fb$^{-1}$) subsample. 
The distributions of a 
different NN output
using a smaller set of input variables and without using the ME discriminant,
trained  on the Run IIa subsamples are shown in~\cite{oldresult}.} 
\label{fig:NNoutput_ee}
\begin{picture}(0,0)(0,0)
  \put (-160,290){\scriptsize {\bf (a)}}
  \put ( 30 ,290){\scriptsize {\bf (b)}}
  \put (-160,170){\scriptsize {\bf (c)}}
  \put ( 30 ,170){\scriptsize {\bf (d)}}
\end{picture}
\end{figure*}

\begin{figure*}[htbp]
\centering
\includegraphics[width=65mm]{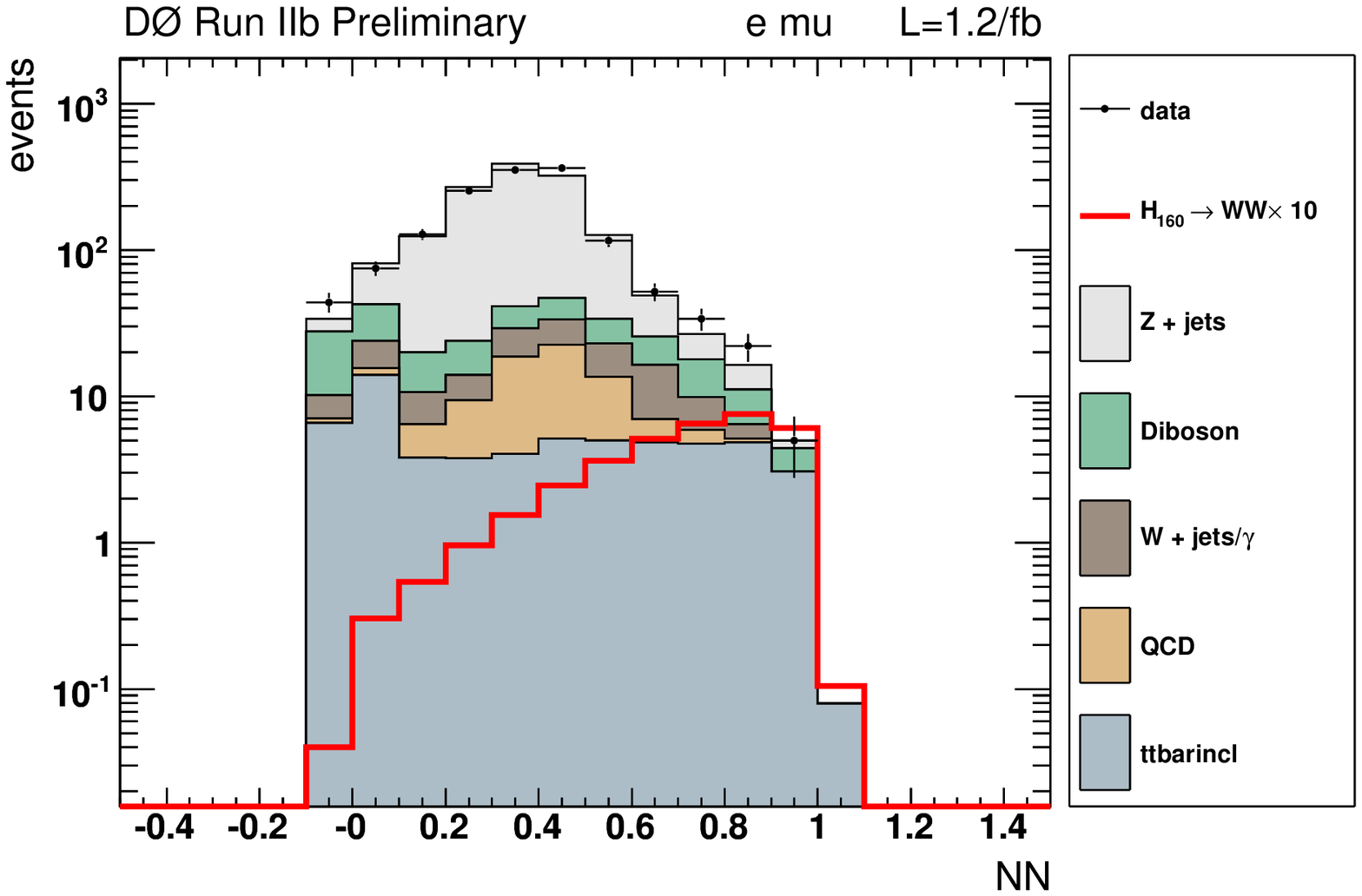}
\includegraphics[width=65mm]{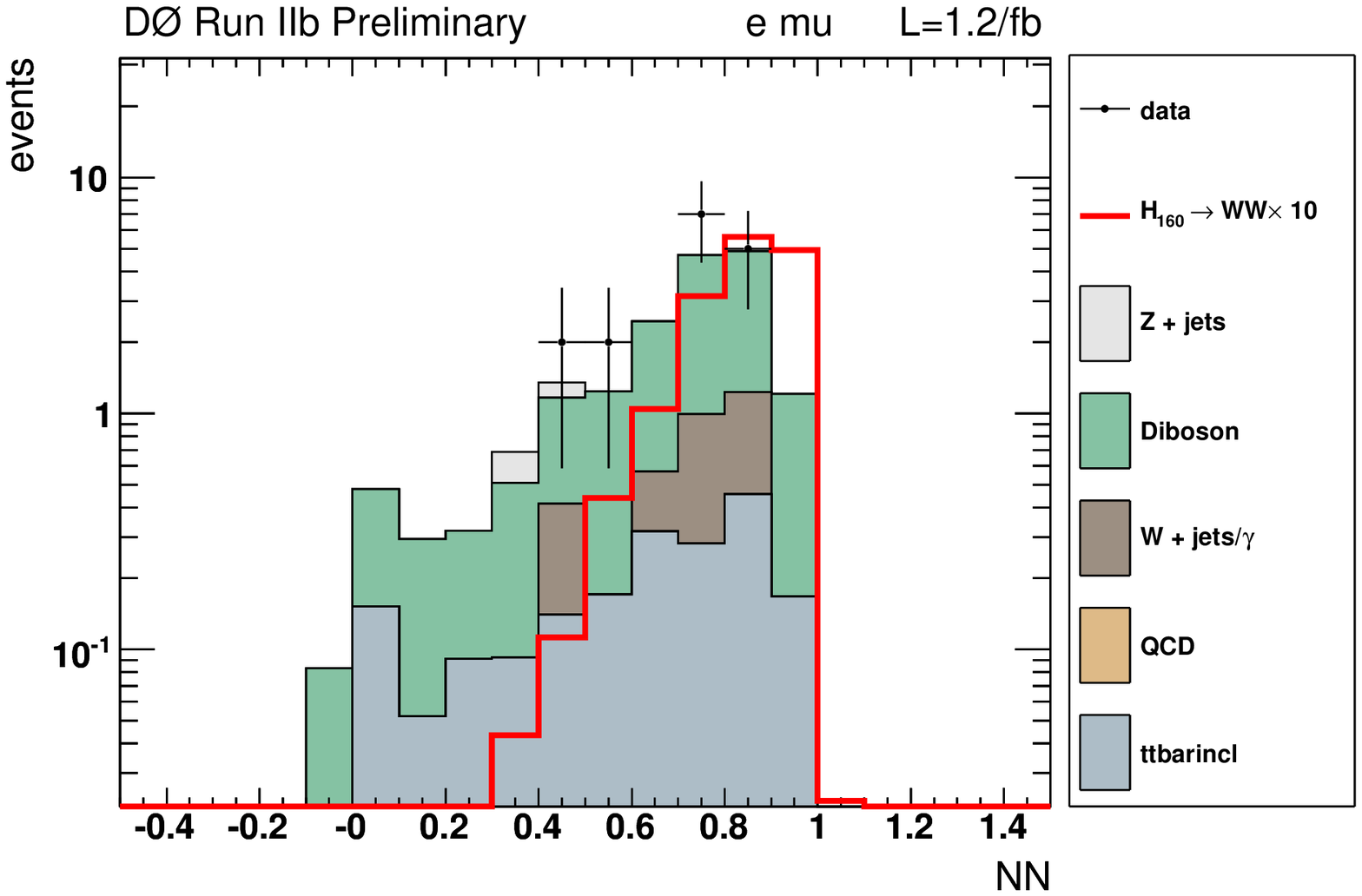}
\caption{Distribution of the NN output in the D\O\ $e\mu$ analysis at the preselection (a) and 
final selection (b) level, in the Run~IIb~(1.2~fb$^{-1}$) subsample. 
For the distributions of the different NN output used in Run IIa, see~\cite{oldresult}.}
\label{fig:NNoutput_emu}
\begin{picture}(0,0)(0,0)
  \put (-160,155){\scriptsize {\bf (a)}}
  \put ( 30,155){\scriptsize {\bf (b)}}
\end{picture}
\end{figure*}

\begin{figure*}[htbp]
\centering
\includegraphics[width=65mm]{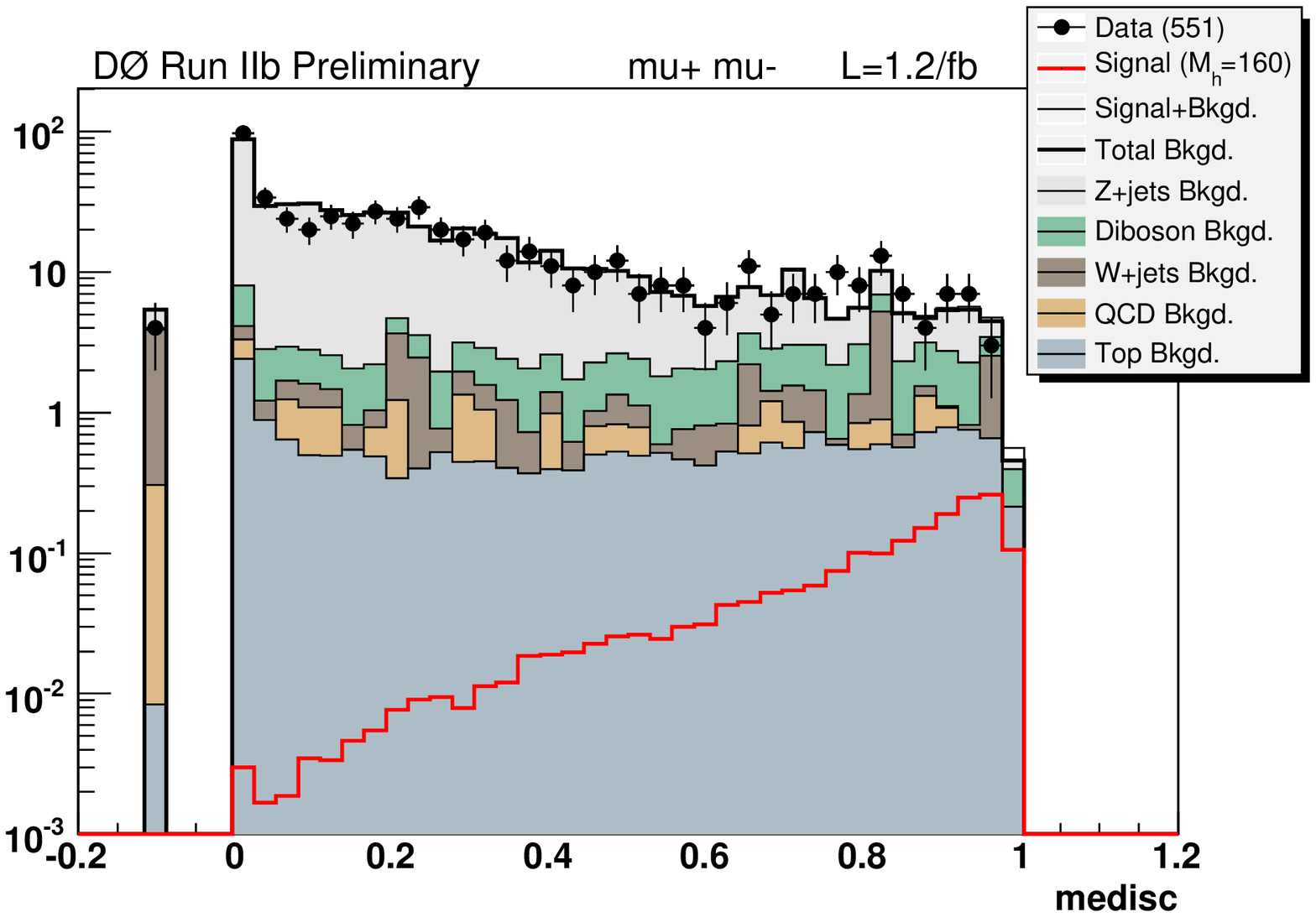}
\includegraphics[width=65mm]{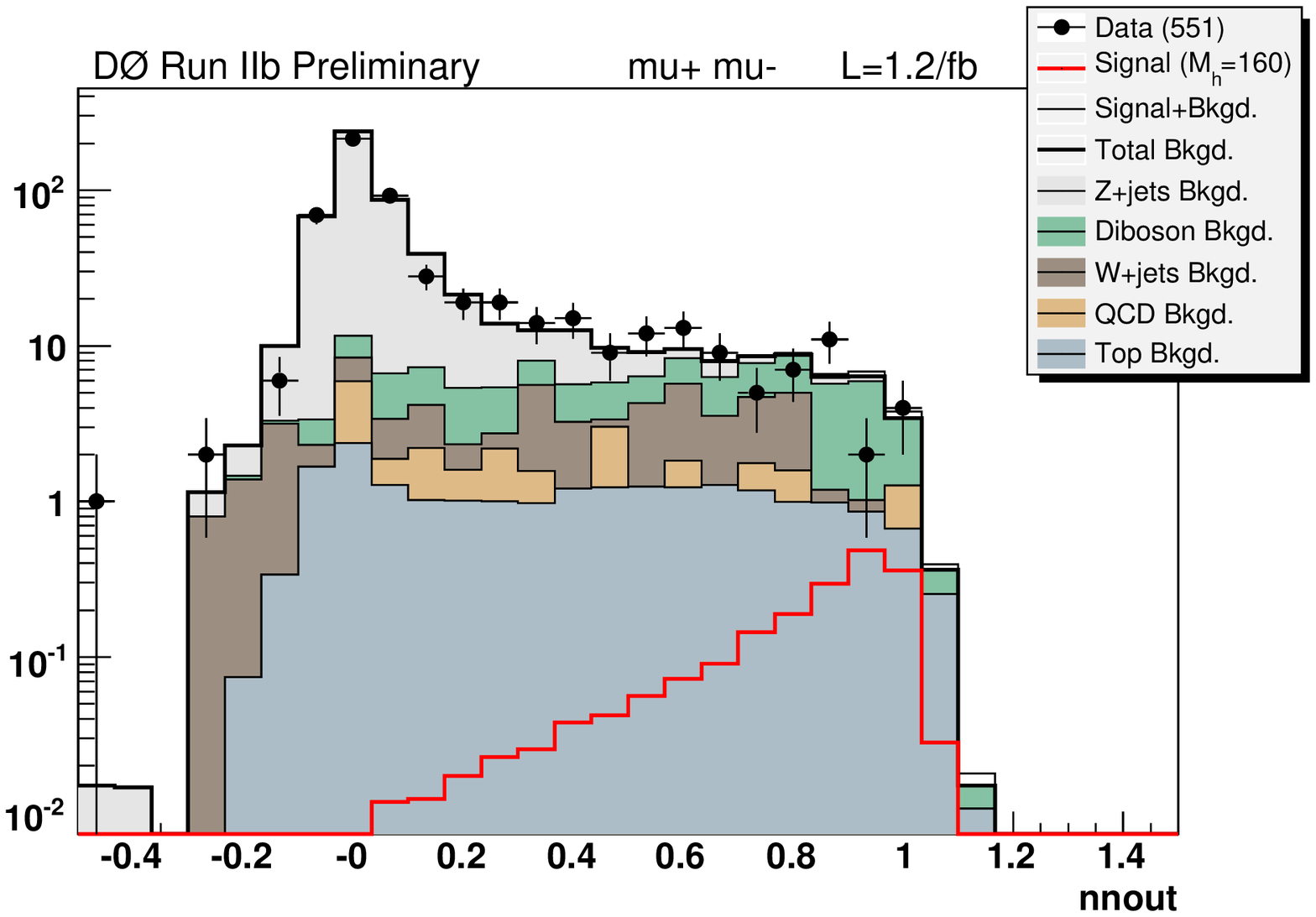}
\caption{Distribution of the ME (a) and NN (b) output in the D\O\ $\mu\mu$ analysis at the 
final selection level, in the Run~IIb~(1.2~fb$^{-1}$) subsample.
For the distributions of the different NN output used in Run IIa, see~\cite{oldresult}.}
\label{fig:NNoutput_mumu}
\begin{picture}(0,0)(0,0)
  \put (-160,155){\scriptsize {\bf (a)}}
  \put ( 30,155){\scriptsize {\bf (b)}}
\end{picture}
\end{figure*}

In the $ee$ and $e\mu$ channels, the signal-to-background is about 10\%. In the $\mu\mu$ channel
it is significantly lower, since the cuts have been relaxed on purpose to increase
the signal acceptance. In these three cases, multivariate techniques are then applied
to enhance the signal.
Neural networks (NN's) are used in each of the three  channels.
and are trained separately for each Higgs
boson mass tested. In the $\mu\mu$ channel, a weighted sum of all
backgrounds is used for the training while for the $ee$ and $e\mu$
channels the NN is trained only against the main $W+jets$ and $WW$
backgrounds.
Input variables to the NN's  are derived based on the separation
power of the various distributions, for each of the three channels.
They  fall into three classes: object
kinematics, event kinematics and angular variables, see~\cite{d0-ww}
for the complete list.
One of the crucial input variables is a discriminant constructed using the 
Matrix Element (ME) method, in which Leading-order parton states 
for either signal ($H
\rightarrow WW$) or $WW$ background are integrated over to find the probability
that the event is signal- or background-like.
The NN  
output 
 for $m_H=160$ GeV is displayed at preselection level
and at final selection level in
Figures~\ref{fig:NNoutput_ee}c,d and \ref{fig:NNoutput_emu}a,b
for the $ee$ and $e\mu$ channels, respectively,
and in Figure~\ref{fig:NNoutput_ee}a,b the corresponding ME
output is shown.
In Figure~\ref{fig:NNoutput_mumu} are shown the ME discriminant and the 
NN output of the $\mu\mu$ channel at the final selection level.

The expected number of background and signal
events depend on efficiencies that lead to the systematic
uncertainties detailed in ~\cite{d0-ww}.
The total uncertainty on the background
 is approximately 16\% and 10\% on  the signal efficiency.
After all selection cuts, the sum of the 
expected backgrounds describe properly the NN output distributions,
which are 
used to set limits on the production cross
section times branching ratio $\sigma \times BR(H \rightarrow W
W^{(*)})$. Limits are determined for each channel
using the CLs method with a log-likelihood ratio
(LLR) test statistic \cite{bib:limits} and are then combined.
 To minimize the 
effects of systematic uncertainties, the
background contributions are fitted to the data observation by
maximizing a profile likelihood function assuming the presence
or the absence of signal
\cite{bib:sys}. 
Table~\ref{tab:alllimitbay} presents the expected and
observed combined 
upper limits at 95\% CL for $\sigma \times BR(H \rightarrow
W W^{(*)})$ relative to that expected in the SM
 for each Higgs boson
mass considered.
Figure~\ref{lim-d0-cdf}a displays the expected and observed limits
for $\sigma\times BR$($H \rightarrow W W^{(*)})$ relative to the SM
for the different Higgs boson masses 
for the current Run II dataset of 2.3 \ifb\ analyzed by \DZ. 

\clearpage

\begin{table}[ht]
\caption{\label{tab:alllimitbay} Expected and observed upper limits
at 95\% CL for $\sigma \times BR (H \rightarrow W W^{(*)}) $
relative to the SM for different Higgs boson masses ($m_H$) as obtained by \DZ.\\}
\begin{tabular}{|c|cccccccccccccccccc|}
\hline
$m_H$ (GeV)& 115& 120& 125& 130& 135& 140& 145& 150& 155& 160& 165& 170& 175& 180& 185& 190& 195& 200 \\
\hline 
Expected  & 48.1 & 16.9 & 12.8 & 8.8 & 7.5 & 6.0 & 5.0 &
4.1 & 3.2 & 2.4 & 2.4 & 2.9 & 3.2 & 3.6 & 4.6 & 5.8 & 6.9 & 8.7 \\
Observed  & 72.4 & 40.8 & 26.1 & 15.7 & 12.3 & 9.9 & 5.5&
4.3 & 3.2 & 2.1 & 2.7 & 2.6 & 3.5 & 3.9 & 3.8 & 4.2 & 7.1 & 6.5 \\
\hline 
\end{tabular}
\end{table}


\subsection{CDF Results on \ \phwww}

In the 2.4 fb$^{-1}$ CDF analysis~\cite{cdf-ww}, the events
are triggered requiring different type of 
electromagnetic (EM) energy cluster for the electron based
samples or track  segments
to record the muon based samples.
Trigger efficiencies
are measured using leptonic W and Z data samples.
To improve the signal acceptance while maintaining
acceptable background rejection for the $W+$ jets and  $W \gamma$
processes where a jet or a $\gamma$ is misidentified as a lepton,
a lepton identification strategy similar to the one 
used in~\cite{cdf-zz}
is used. Candidate
leptons are separated into six orthogonal sets:
two for electrons; three for muons; and one
for tracks that extrapolate outward to detector regions
with insufficient calorimeter coverage for energy measurement.
The two electron sets are central ($|\eta| <$ 1.1)
using drift-chamber-based tracking,  and forward
($1.2 < |\eta| < 2.0$) using  silicon-detector-based 
tracking. One of the muon set uses the
muon chambers while the  other two use tracks matched with
energy deposits consistent with minimum ionization in
the central or forward calorimeters.
All lepton candidates are required to be isolated~\cite{cdf-ww}.
Candidate events are required to 
have exactly two lepton candidates. At least one lepton
must match a trigger lepton candidate and have
$p_T >$ 20 GeV, while the other one must have $p_T >$ 10 GeV.

Events are simulated with the \textsc{mc@nlo} program
for WW~\cite{mcnlo}, \textsc{Pythia} for the signal, Drell Yan, $WZ$, $ZZ$, and
$t\bar{t}$, and the generator described in~\cite{wgam} for
$W\gamma$.
The diboson backgrounds are suppressed in a similar way as in the
 \DZ analysis, with a specific treatment of the \etmiss~\cite{cdf-ww},
while the suppression of the $t\bar{t}$ contribution is achieved
by requiring fewer than two reconstructed
jets with $E_T >$ 15 GeV and $|\eta| <$ 2.5 in the event. The $W\gamma$
and $W+$ jet backgrouns rejections are estimated from data.
 
After selection 
661 events are observed while
626 $\pm$ 54 background events are expected.
The samples are then divided into high and low S/B classes. 
The dominant remaining bacgkround is $WW$ production, and to separate it
from the $H \rightarrow WW$ signal two different multivariate techniques
are combined, as in \DZ, i.e. a neural network approach using also  Matrix
Element discriminants, built as a Likelihood ratio 
($LR_{signal}$ of the event probabilities
$P_{signal}/[P_{signal}+\Sigma_i P_{background_i}]$.
However, while \DZ is using only one $LR$ with
one signal ($H \rightarrow WW$) and one main background ($WW$ production), 
CDF builds several $LR$, assuming 
successively each important background as "signal". The final NN uses then six kinematic
input variables, as shown in Figure~\ref{NN-cdf} and five $LR's$ as shown in
Figure~\ref{ME-cdf}. The resulting NN output is shown in
Figure~\ref{ME-cdf}f for the high S/B case.

Since no excess is observed,  limits are obtained using a Bayesian technique
from the NN outputs, taking into account systematics uncertainties
and their correlations as described in~\cite{cdf-ww}.
The resulting limits are given in Table~\ref{cdf-limit}.
These limits are compared to the ones obtained by \DZ in Figure~\ref{lim-d0-cdf}.
\begin{table}[ht]
\caption{\label{cdf-limit}Expected and observed upper limits
at 95\% CL for $\sigma \times BR (H \rightarrow W W^{(*)}) $
relative to the SM for different Higgs boson masses ($m_H$) as obtained by CDF.\\}
\begin{tabular}{|c|cccccccccc|}
\hline
$m_H$ (GeV)   & 110  & 120  & 130  &  140&  150&  160&  170&  180&  190&  200 \\
\hline 
Expected      & 59.6 & 19.1 &  9.2 & 5.8 & 4.2 & 2.5 & 2.7 & 3.9 & 6.1 &  8.3 \\
Observed      & 53.4 & 15.8 &  5.3 & 3.2 & 2.4 & 1.6 & 1.8 & 2.8 & 5.2 & 10.0 \\
\hline 
\end{tabular}
\end{table}

\begin{figure*}[htbp]
\centering
\includegraphics[height=30mm,width=55mm]{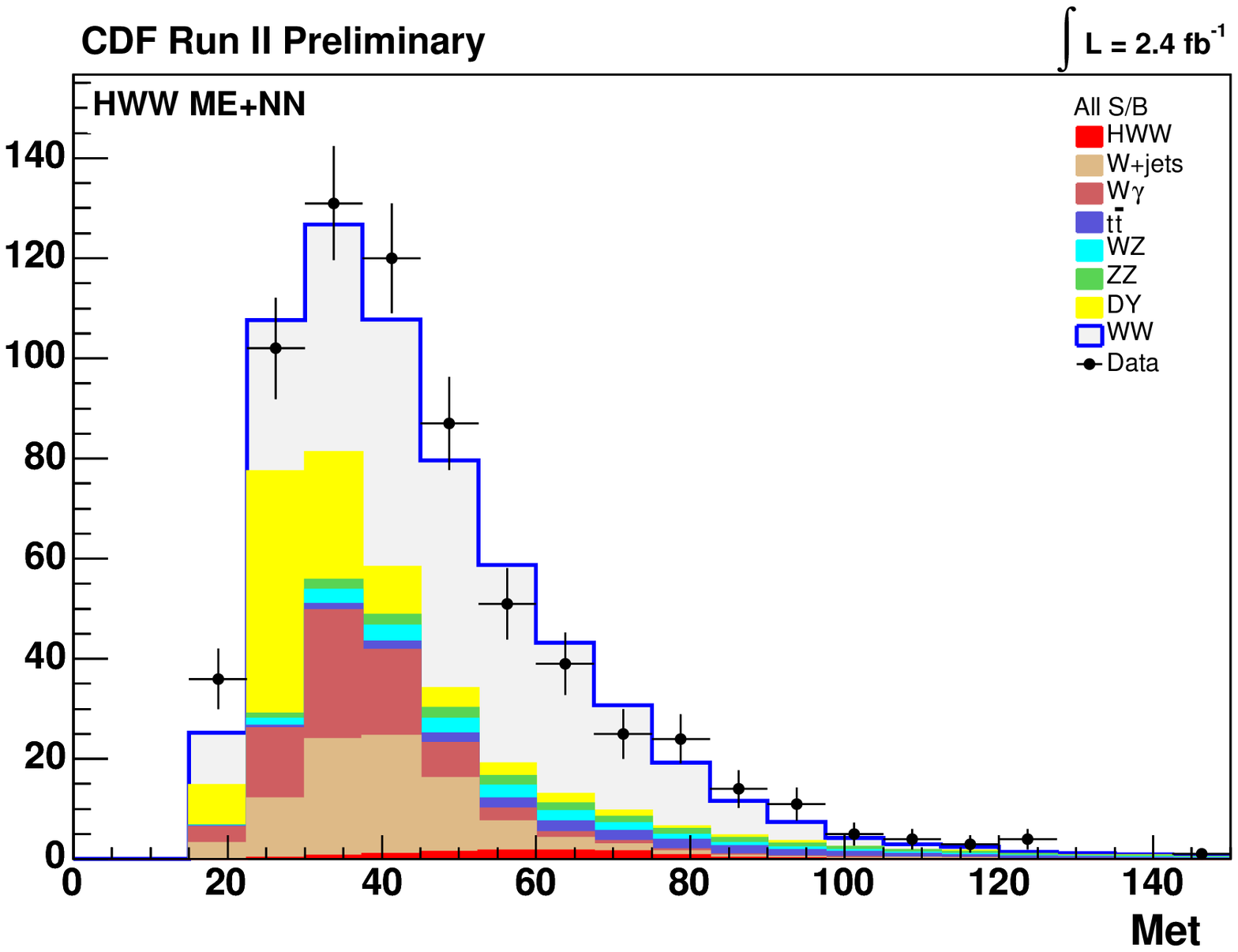}
\includegraphics[height=30mm,width=55mm]{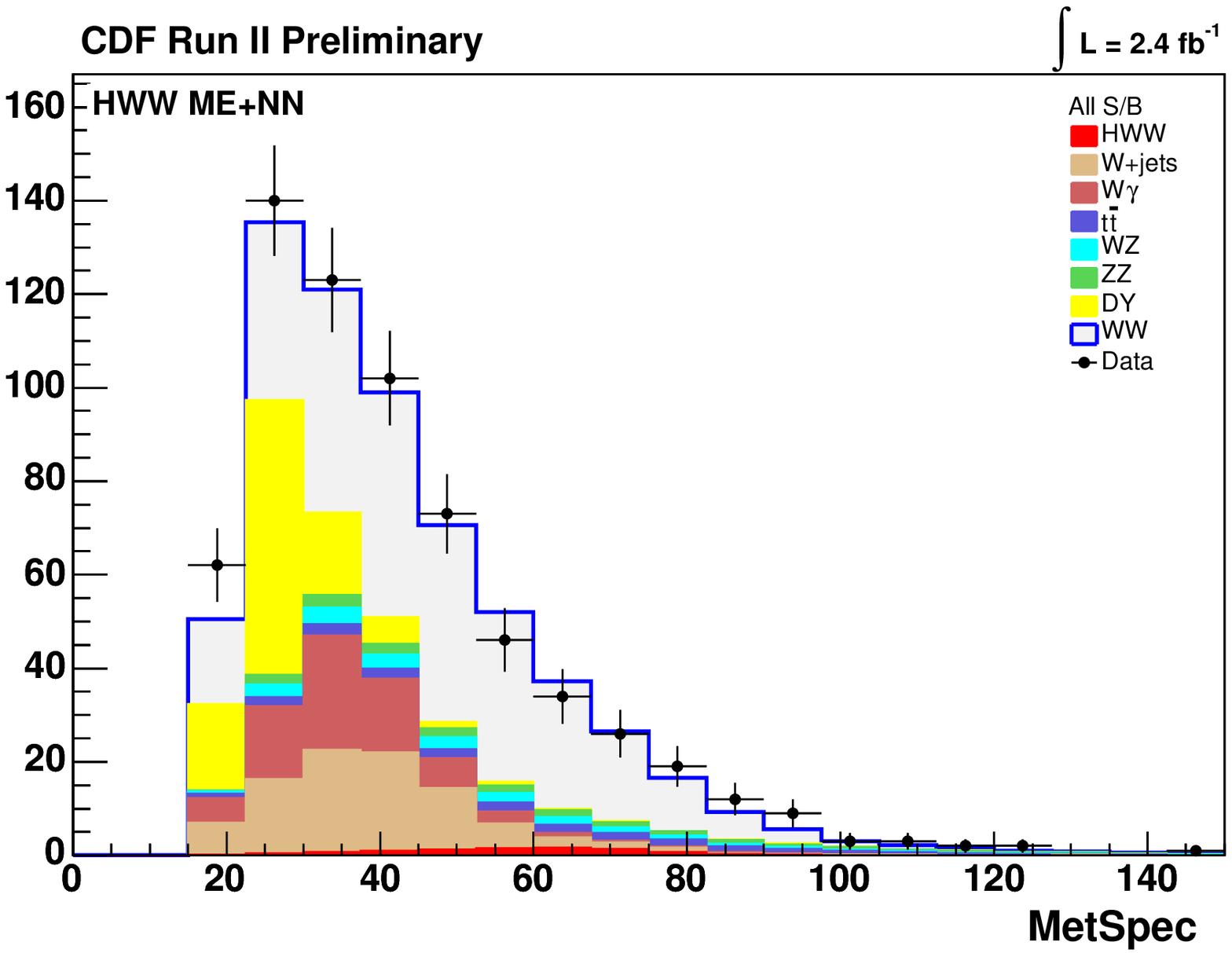}
\includegraphics[height=30mm,width=55mm]{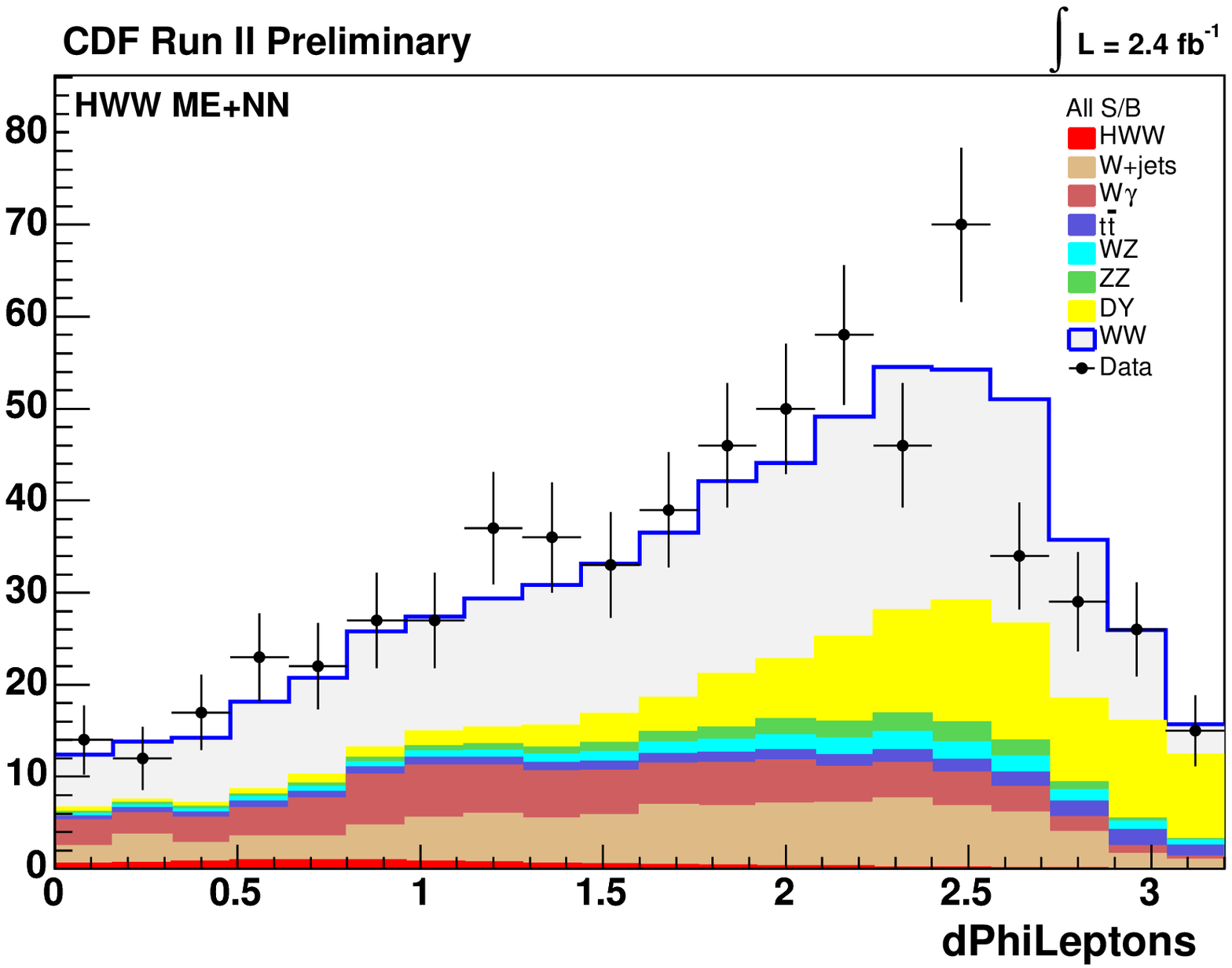}
\includegraphics[height=30mm,width=55mm]{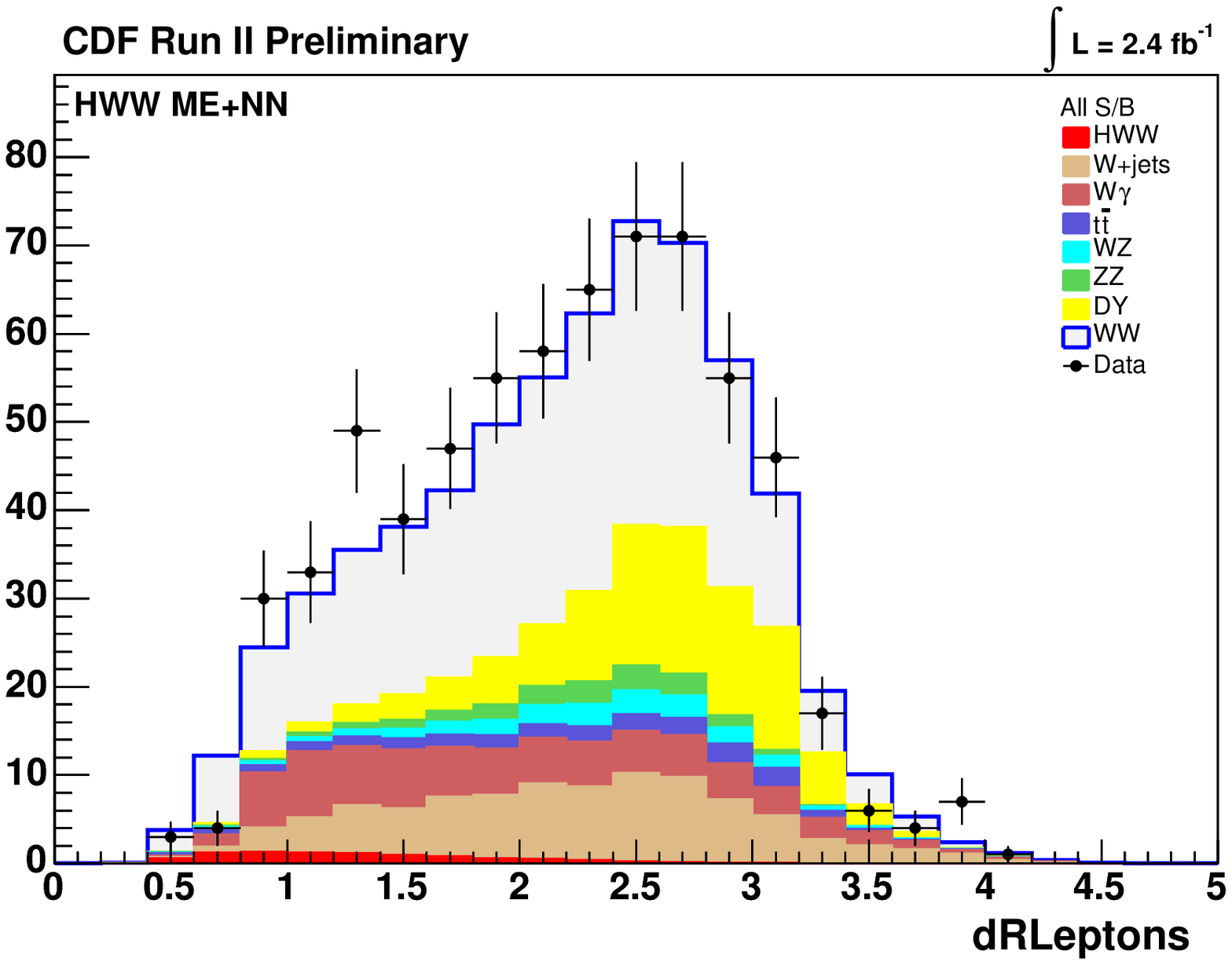}
\includegraphics[height=30mm,width=55mm]{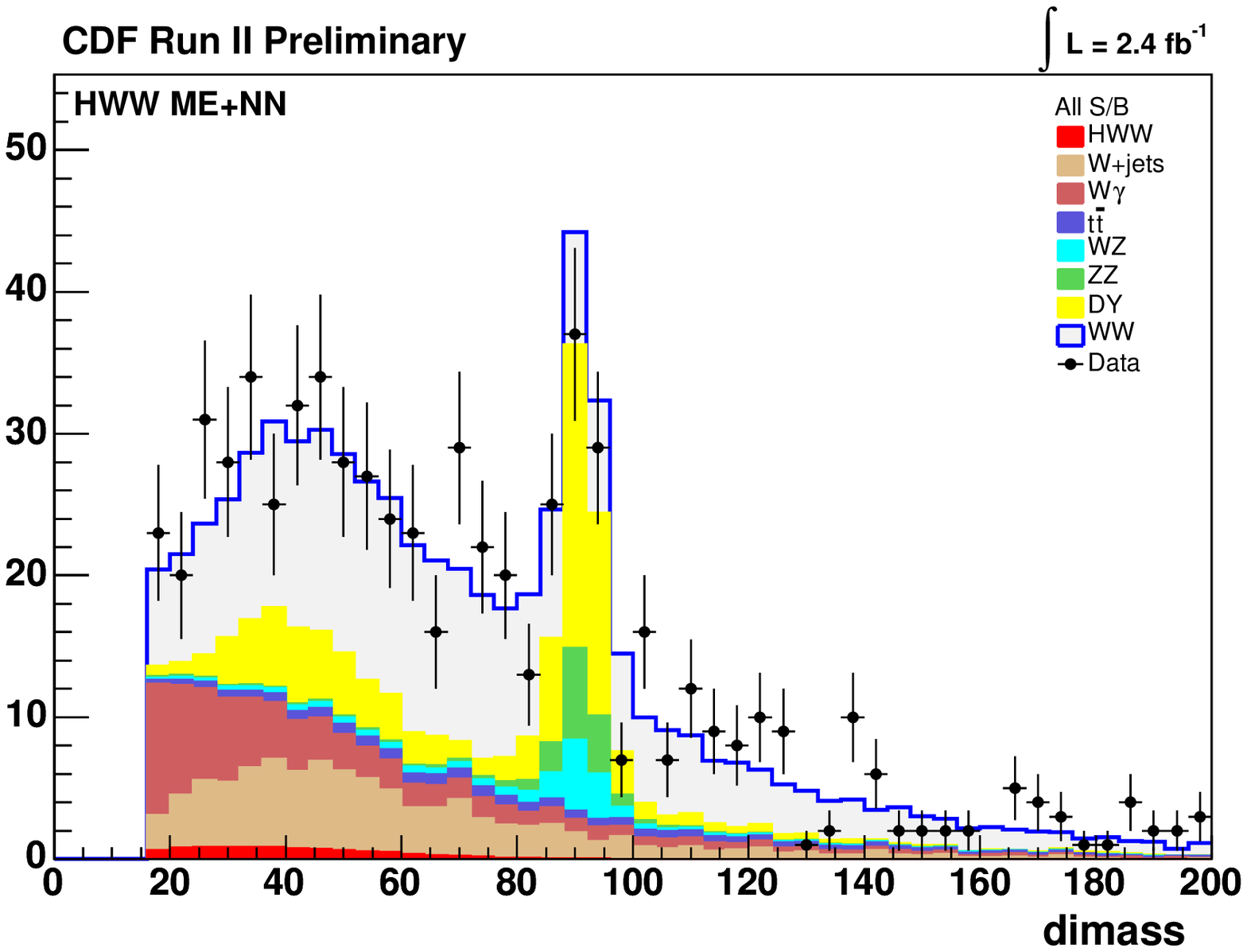}
\includegraphics[height=30mm,width=55mm]{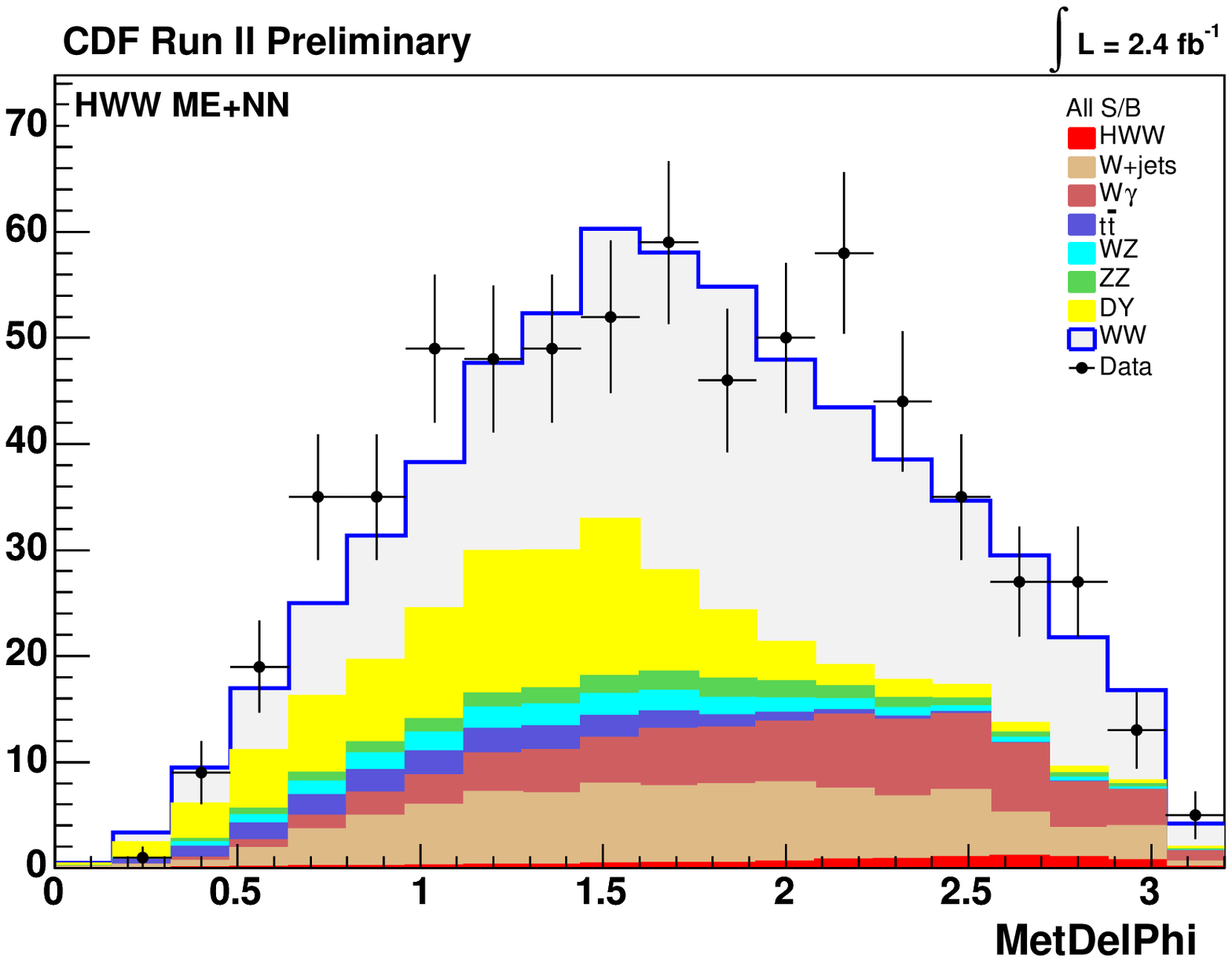}
\caption{(a)-(e) Distributions of the six kinematic variables used 
as input to the NN of the
CDF analysis. The variable definitions are given in ~\cite{cdf-ww}} 
\label{NN-cdf}
\begin{picture}(0,0)(0,0)
  \put (-160,205){\scriptsize {\bf (a)}}
  \put (   0,205){\scriptsize {\bf (b)}}
  \put ( 160,205){\scriptsize {\bf (c)}}
  \put (-160,120){\scriptsize {\bf (d)}}
  \put (   0,120){\scriptsize {\bf (e)}}
  \put ( 160,120){\scriptsize {\bf (f)}}
\end{picture}
\end{figure*}
\begin{figure*}[htbp]
\centering
\includegraphics[height=30mm,width=55mm]{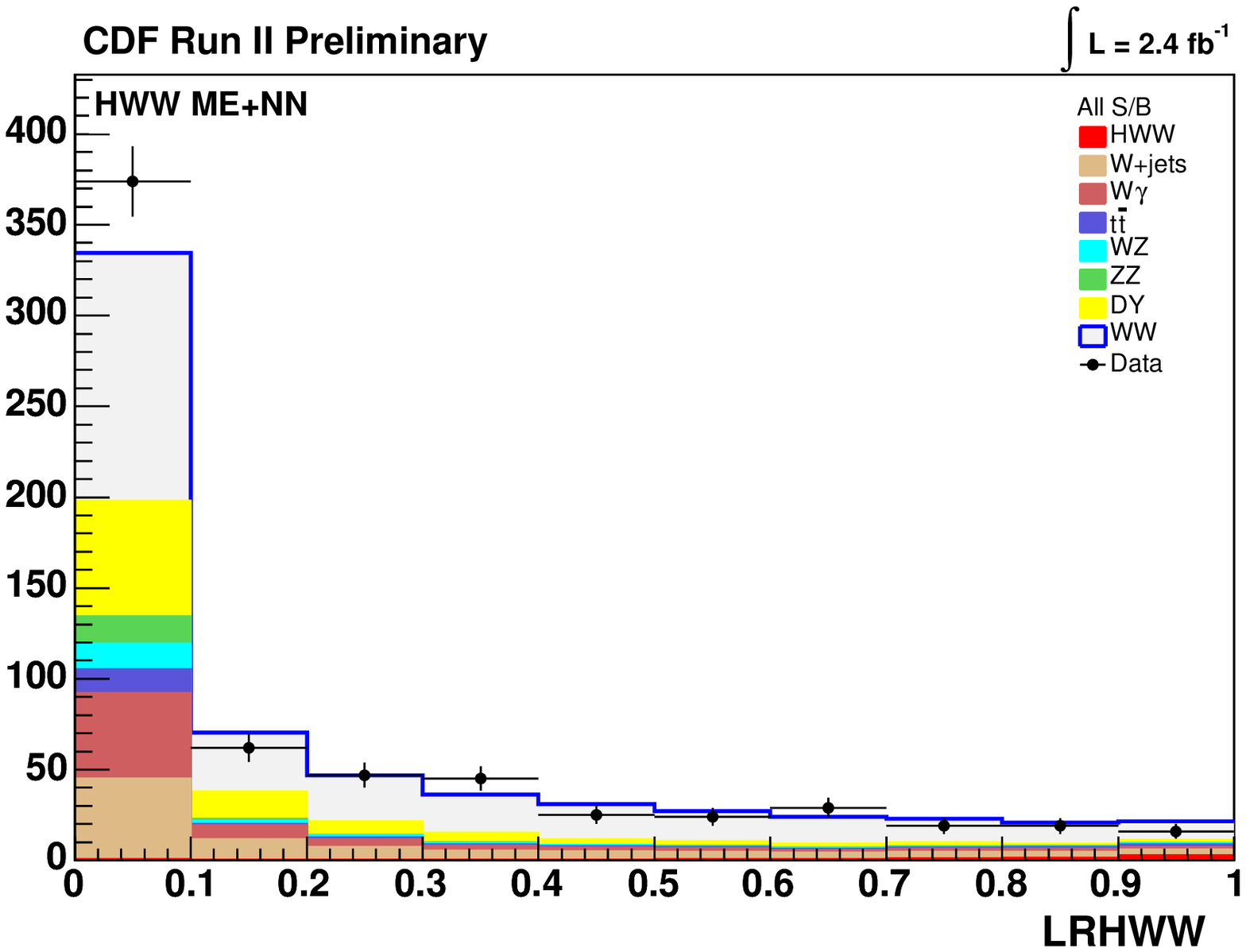}
\includegraphics[height=30mm,width=55mm]{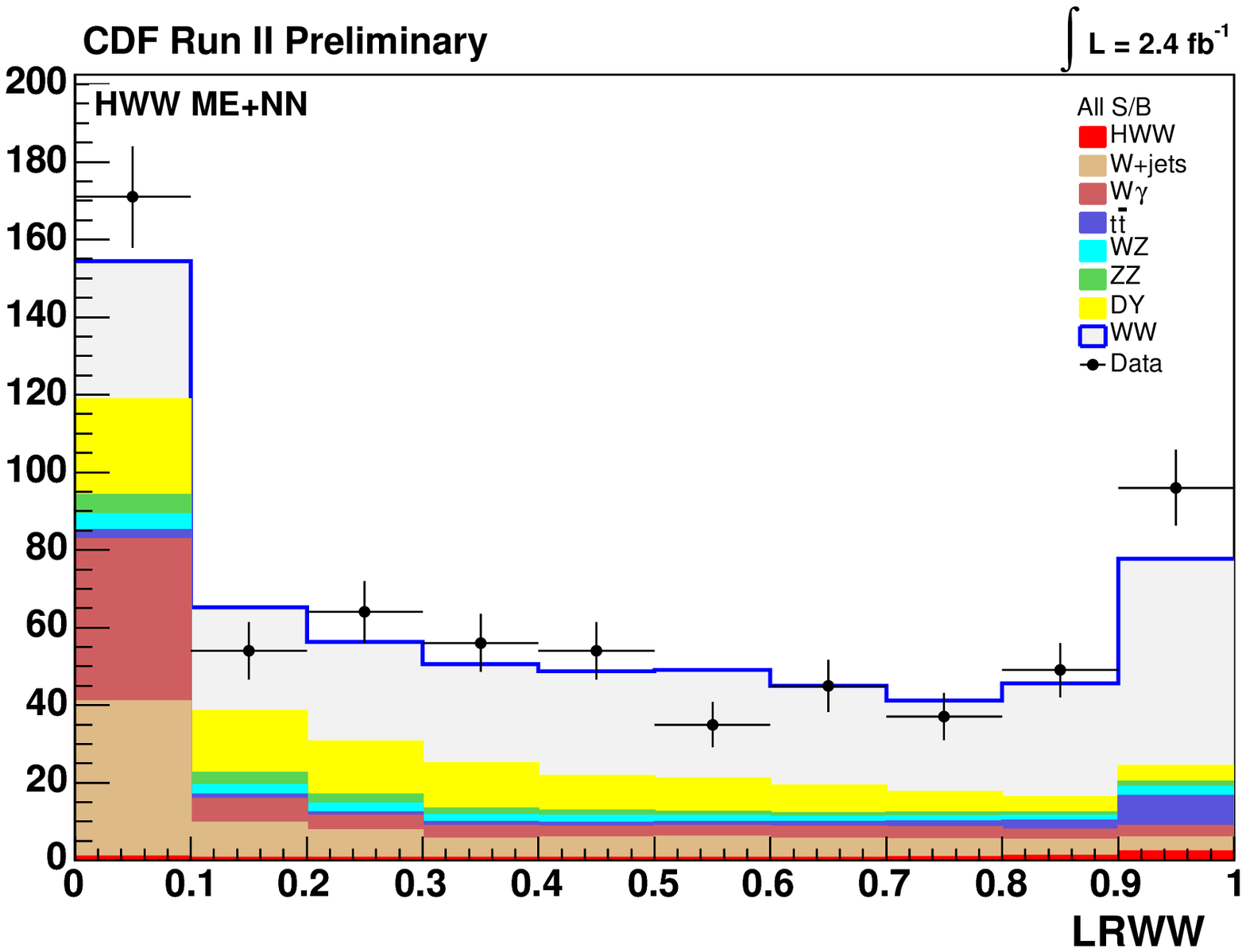}
\includegraphics[height=30mm,width=55mm]{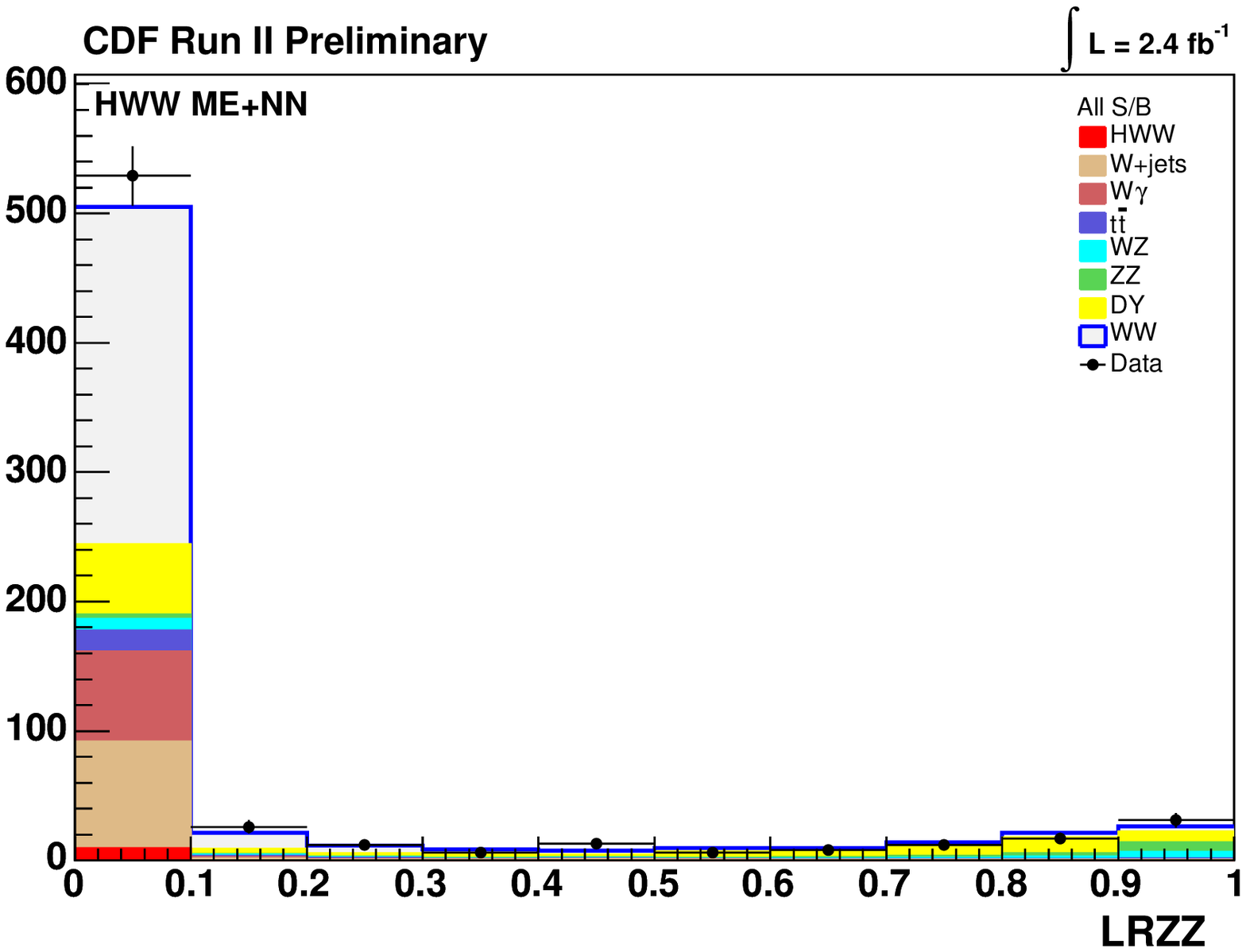}
\includegraphics[height=30mm,width=55mm]{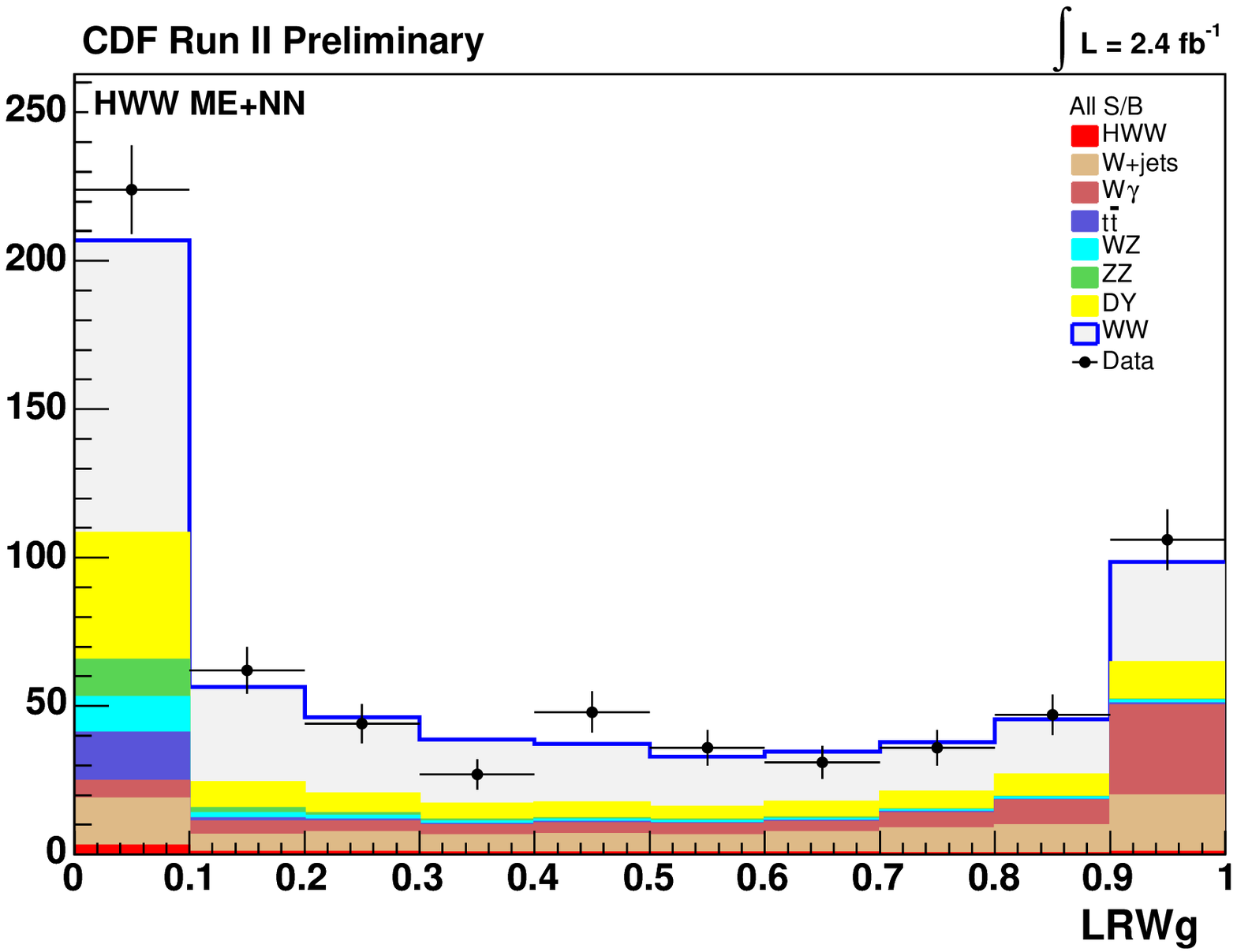}
\includegraphics[height=30mm,width=55mm]{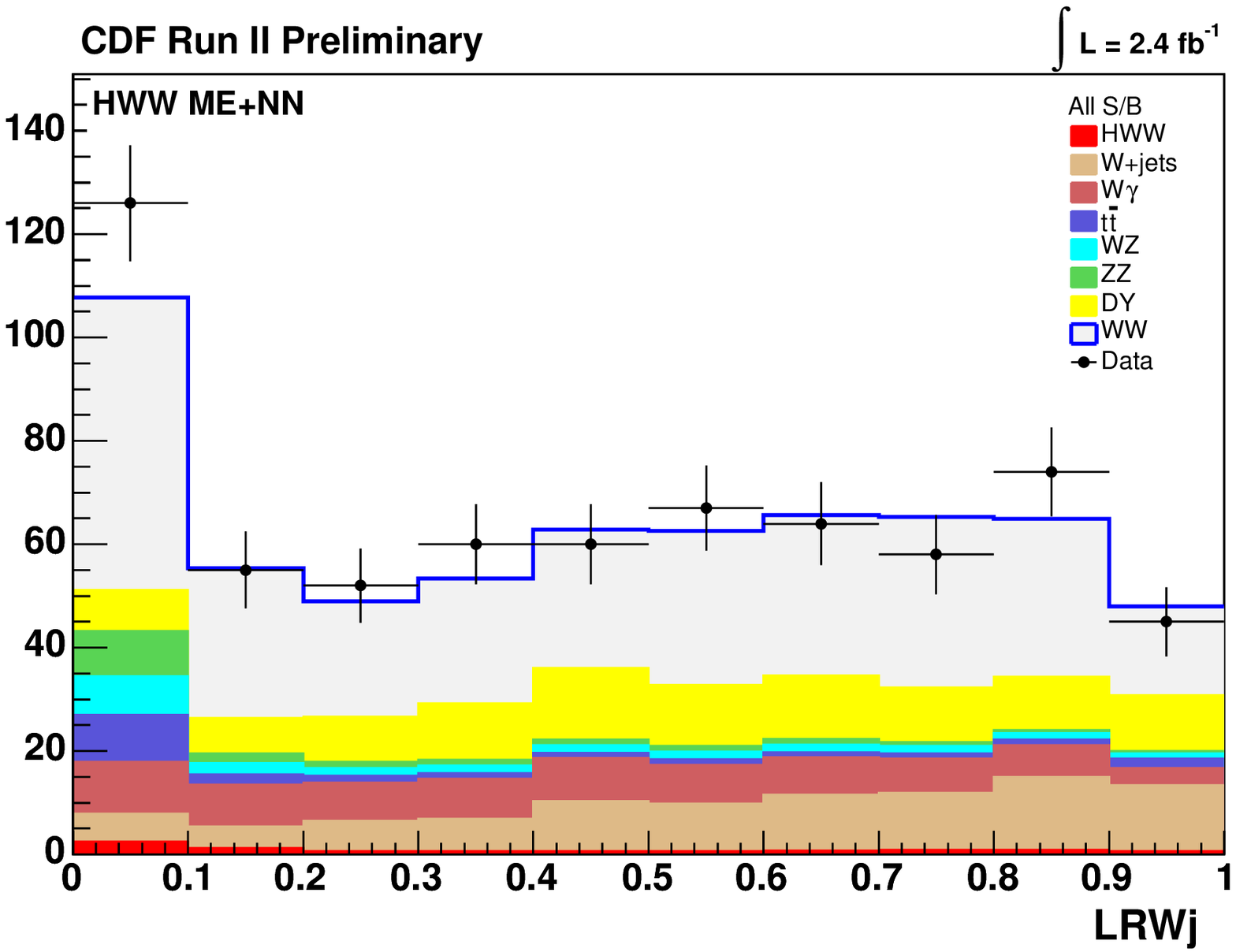}
\includegraphics[height=30mm,width=55mm]{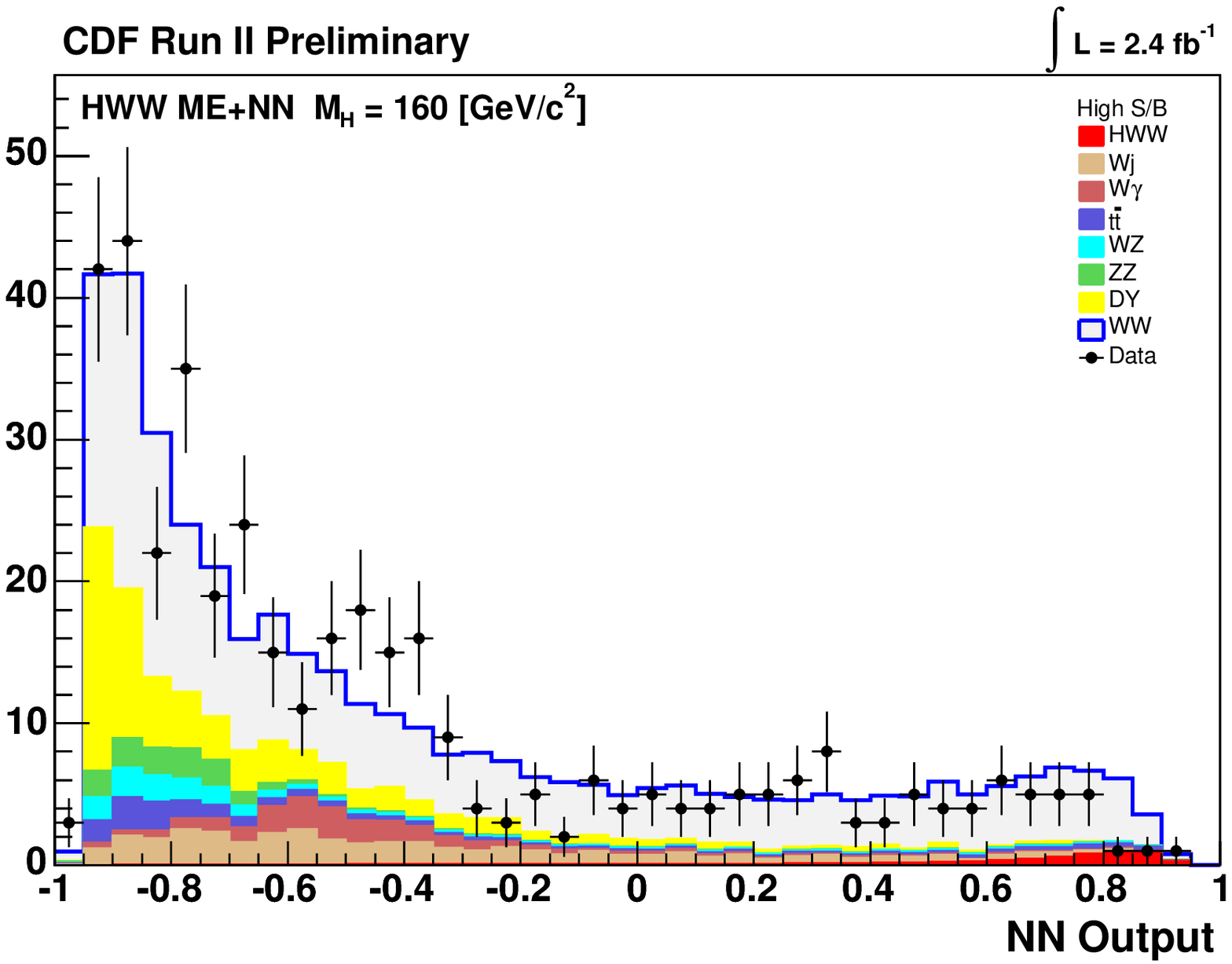}
\caption{(a)-(e) Distributions of the Likelihood ratios based on ME calculations
 used as input to the NN of the
CDF analysis. Besides $LR_{H\rightarrow WW}$, the other $LR's$ are calculated assuming
each background as ``signal''.
(f) Resulting NN output for the high S/B sample}
\label{ME-cdf}
\begin{picture}(0,0)(0,0)
  \put (-160,215){\scriptsize {\bf (a)}}
  \put (   0,215){\scriptsize {\bf (b)}}
  \put ( 160,215){\scriptsize {\bf (c)}}
  \put (-160,130){\scriptsize {\bf (d)}}
  \put (   0,130){\scriptsize {\bf (e)}}
  \put ( 160,130){\scriptsize {\bf (f)}}
\end{picture}
\end{figure*}
\begin{figure*}[htbp]
\centering
\includegraphics[width=70mm]{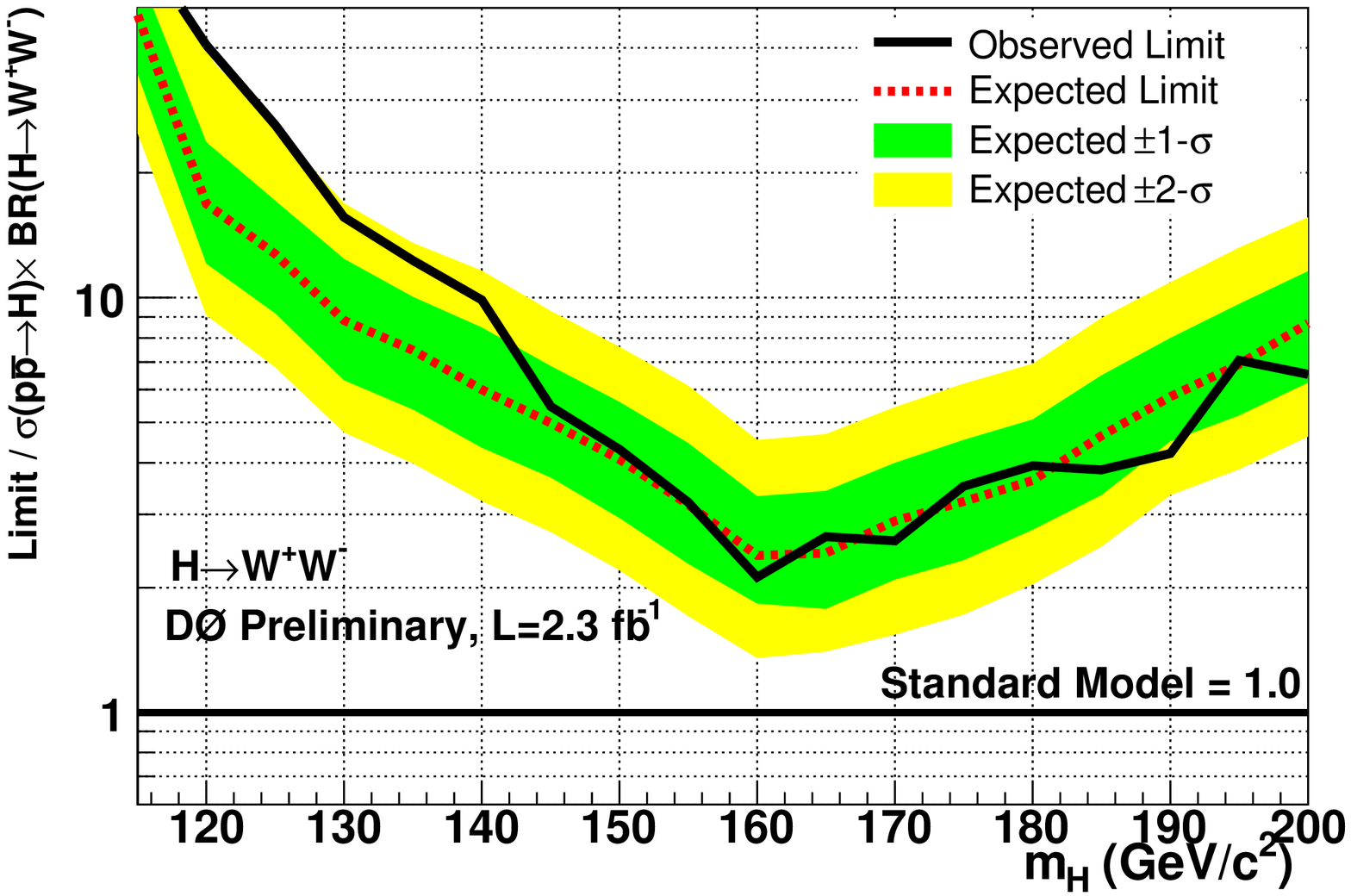}
\includegraphics[width=72mm]{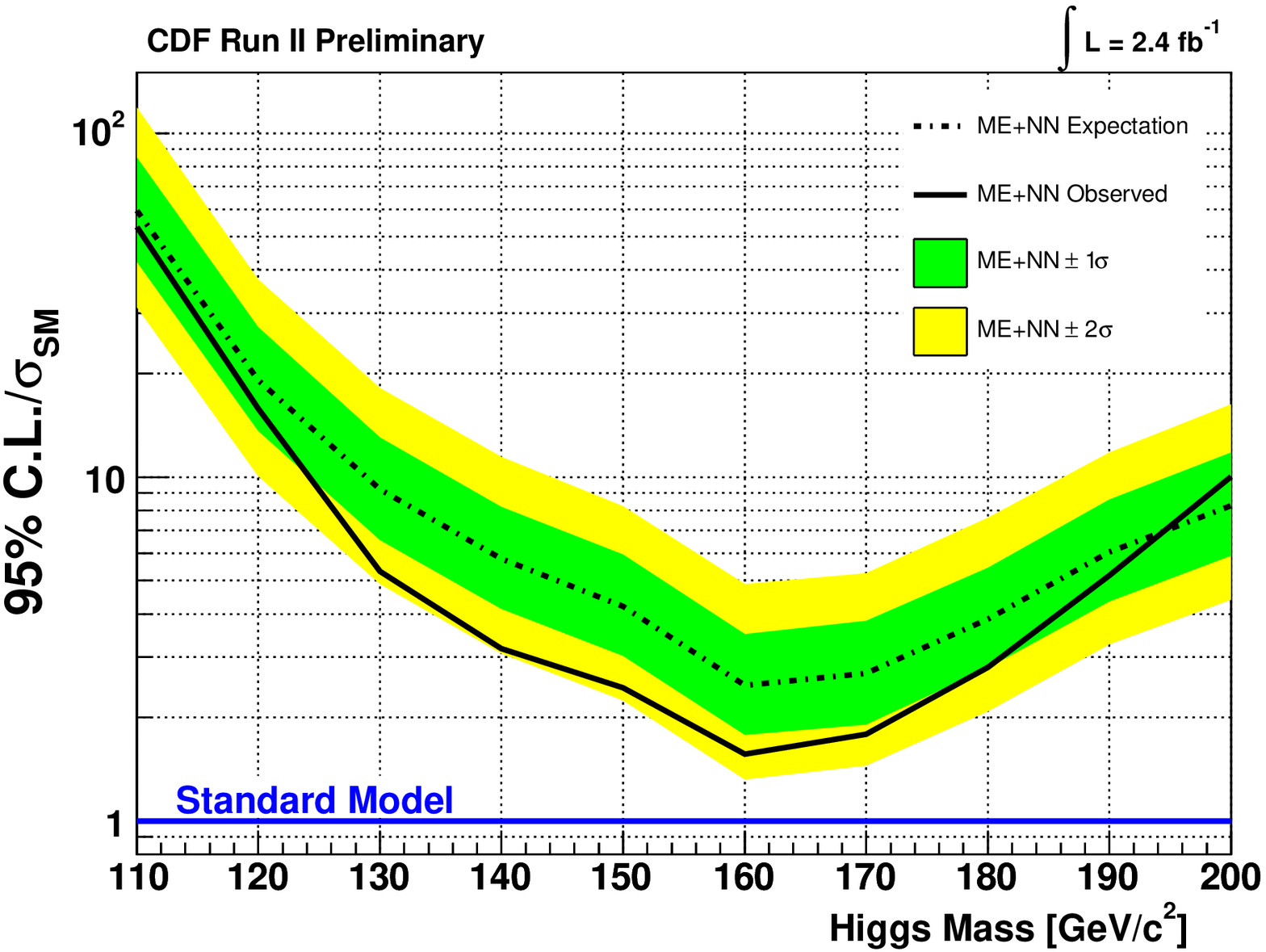}
\caption{\DZ (a) and CDF (b) ratios of 95\% C.L. limits on SM Higgs production
divided by the SM expected cross sections, as a function of $m_H$. An observed
 ratio of 1
is equivalent to  a 95\% C.L. exclusion.}
\label{lim-d0-cdf}
\begin{picture}(0,0)(0,0)
  \put (-145,165){\scriptsize {\bf (a)}}
  \put ( 40,165){\scriptsize {\bf (b)}}
\end{picture}
\end{figure*}

\clearpage

%
%
%
%
%
%

\section{COMBINATION OF CDF AND \DZ RESULTS}

CDF and \DZ have combined their most recent 
results on the  searches for a SM Higgs
boson produced in association with vector bosons 
(\pwh, \pzh~or \pwww) or
through gluon-gluon fusion (\phww) or vector boson fusion (VBF), in data
corresponding to integrated luminosities ranging from 1.0-2.4\ifb~at
CDF and 1.1-2.3\ifb~at D\O . 
In this combination the searches for Higgs bosons
decaying to two photons or two tau leptons are added for the first time.
The searches are separated into twenty nine
mutually exclusive 
final states~\cite{cdf-d0-comb} 
referred to as ``analyses'' in the following.
%
Integrated luminosities, 
and references to the collaborations' public documentation for each analysis
are given in Table~\ref{tab:cdfacc}
for CDF and in Table~\ref{tab:dzacc} for D\O .  
The tables include the ranges of Higgs boson mass ($m_H$) over which
the searches were performed. 

\begin{table}[h]
\caption{\label{tab:cdfacc}Luminosity, explored mass range and references 
for the CDF analyses.  $\ell$ stands for either $e$ or $\mu$.
}
\begin{tabular}{|lccccc|}
\hline
&$WH\rightarrow \ell\nu b\bar{b}$ & $ZH\rightarrow \nu\bar{\nu} b\bar{b}$ &
$ZH\rightarrow \ell^+\ell^- b\bar{b}$ &  $H\rightarrow W^+ W^- $ &   $H$ + $X
\rightarrow \tau^+ \tau^- $ + 2 jets \\ 
\hline 
Luminosity (\ifb)         & 1.9        &  1.7            & 1.0      & 2.4     & 2.0 \\ 
$m_{H}$ range (GeV/c$^2$) & 110-150    & 100-150         & 110-150  & 110-200 & 110-150  \\
Reference       & \cite{cdfWH} & \cite{cdfZH}& \cite{cdfZHll} & \cite{cdfHWW} & \cite{cdfHtt}   \\
\hline
\end{tabular}
\end{table}
\vglue -1cm 
\begin{table}[h]
\caption{\label{tab:dzacc}Luminosity, explored mass range and references 
for the D\O\ analyses.  $\ell$ stands for either $e$ or $\mu$.
}
\begin{tabular}{|lcccccc|}
\hline
&$WH\rightarrow \ell\nu b\bar{b}$ & $ZH\rightarrow \nu\bar{\nu} b\bar{b}$ &
$ZH\rightarrow \ell^+\ell^- b\bar{b}$ &  $H\rightarrow W^+ W^- $ & 
 $WH \rightarrow WW^+ W^-$ & $H \rightarrow \gamma \gamma$ \\ 
\hline 
Luminosity (\ifb)         & 1.7        &  2.1            & 1.1 & 2.3 & 1.1 & 2.3\\ 
$m_{H}$ range (GeV) & 105-145    & 105-145         & 105-145  & 110-200 & 120-200 & 105-145\\
Reference       & \cite{dzWHl} & \cite{dzZHv} & \cite{dzZHll} 
& \cite{dzHWW},\cite{dzHWW-2b} 
 & \cite{dzWWW}  & \cite{dzHgg} 
\\
\hline
\end{tabular}
\end{table}

Several types of combinations, using the
Bayesian and  Modified Frequentist approaches, have been performed
to ensure that the final result does not depend on the details 
of the statistical approach, and indeed, similar results
(within 10\%) have been obtained.
Both
methods rely on distributions in the final discriminants, and not just on
their single integrated
values.  Systematic uncertainties enter as uncertainties on the
expected number of signal and background events, as well
as on the distribution of the discriminants in 
each analysis (``shape uncertainties'').
Both methods use likelihood calculations based on Poisson
probabilities. 
The Bayesian Method and Modified Frequentist method used are described
in~\cite{cdf-d0-comb}.

\subsection{Systematic Uncertainties} 

Systematic uncertainties differ
between experiments and analyses, and they affect the rates and shapes of the predicted
signal and background in correlated ways.  The combined results incorporate
the sensitivity of predictions to  values of nuisance parameters,
and correlations are included, between rates and shapes, between signals and backgrounds,
and between channels within experiments and between experiments.
More on these issues can be found  in~\cite{cdf-d0-comb} and in the
individual analysis notes~\cite{cdfWH}-\cite{dzHgg}.
The main uncertainties are listed here below:
\begin{itemize} 
\item {\bf Correlated Systematics between CDF and \DZ:}
~of the 6\% uncertainty on the measurement of the integrated luminosity obtained
by each experiment,
4\% arises from the uncertainty
on the inelastic \pp~scattering cross section, which is correlated
between CDF and D\O . 
The uncertainty on the production rates for the signal, for 
top-quark processes (\ttbar~and single top) and for electroweak processes
($WW$, $WZ$, and $ZZ$) are taken as correlated between the two
experiments. 
The 
uncertainties on the background rates for $W/Z$+heavy flavor 
are considered uncorrelated, as both CDF and D\O\ estimate
these rates using data control samples, but employ different techniques.
Other data driven uncertainty determinations (multijet, fake lepton or $b$-id rates)
are  taken uncorrelated between the two experiments for the same reason.
\item{\bf Correlated Systematic Uncertainties for CDF:}
~for \hbb, the largest
uncertainties on signal arise from a scale factor for $b$-tagging 
(5.3-16\%), jet energy scale (1-20\%) and MC modeling (2-10\%). 
The shape dependence of the jet energy scale, $b$-tagging and
uncertainties  on gluon radiation 
are taken into account for some analyses.
For
\hww, the largest uncertainty comes from
MC modeling (5\%).  For simulated backgrounds, the uncertainties on the
expected rates range from 11-40\%, depending on the background.
\item{\bf Correlated Systematic Uncertainties for D\O :}
~\hbb~analyses have an uncertainty on the
$b$-tagging rate of 3-10\% per jet, and  also an
uncertainty on the jet energy and acceptance of 6-9\% (jet
identification, energy calibration and resolution).
For the high mass analyses, the largest uncertainties are associated with lepton
measurement and acceptance. These values range from 2-11\% depending on
the final state.  The largest contributing factor to all analyses is
the uncertainty on cross sections for simulated background, and is 
6-18\%. 
\end{itemize}

\subsection{Combined Results} 

Before extracting the combined limits we study the distributions 
of the 
log-likelihood ratio (LLR) for different hypothesis,
to check the expected
sensitivity across the mass range tested.
Figure~\ref{combos}a
displays the LLR distributions
for the combined
analyses as a function of $m_{H}$. Included are the results for the
background-only hypothesis (LLR$_{b}$), the signal and background
hypothesis (LLR$_{s+b}$), and for the data (LLR$_{obs}$).  The
shaded bands represent the 1 and 2 standard deviation ($\sigma$)
departures for LLR$_{b}$. 


Using the combination procedures outlined in~\cite{cdf-d0-comb}, we extract limits on
SM Higgs boson production $\sigma \times B(H\rightarrow X)$ in
\pp~collisions at $\sqrt{s}=1.96$~TeV. 
For a simpler comparison with the standard model 
 we present our results in terms of
the ratio of obtained limits  to  cross section in the SM, as a function of
Higgs boson mass, for test masses for which
both experiments have performed dedicated searches in different channels.
 A value of 1 would indicate a Higgs boson mass excluded at
95\% C.L. The expected and observed 95\% C.L. ratios to the
SM cross section for the combined CDF and D\O\ analyses are shown in
Figure~\ref{combos}b.  The observed and median expected limit ratios
are given  in Table~\ref{tab:ratios}, with
observed (expected) values of 
3.7 (3.3) at $m_{H}=115$~GeV and
1.1 (1.6) at $m_{H}=160$~GeV.
\begin{figure*}[b]
\centering
\includegraphics[width=75mm]{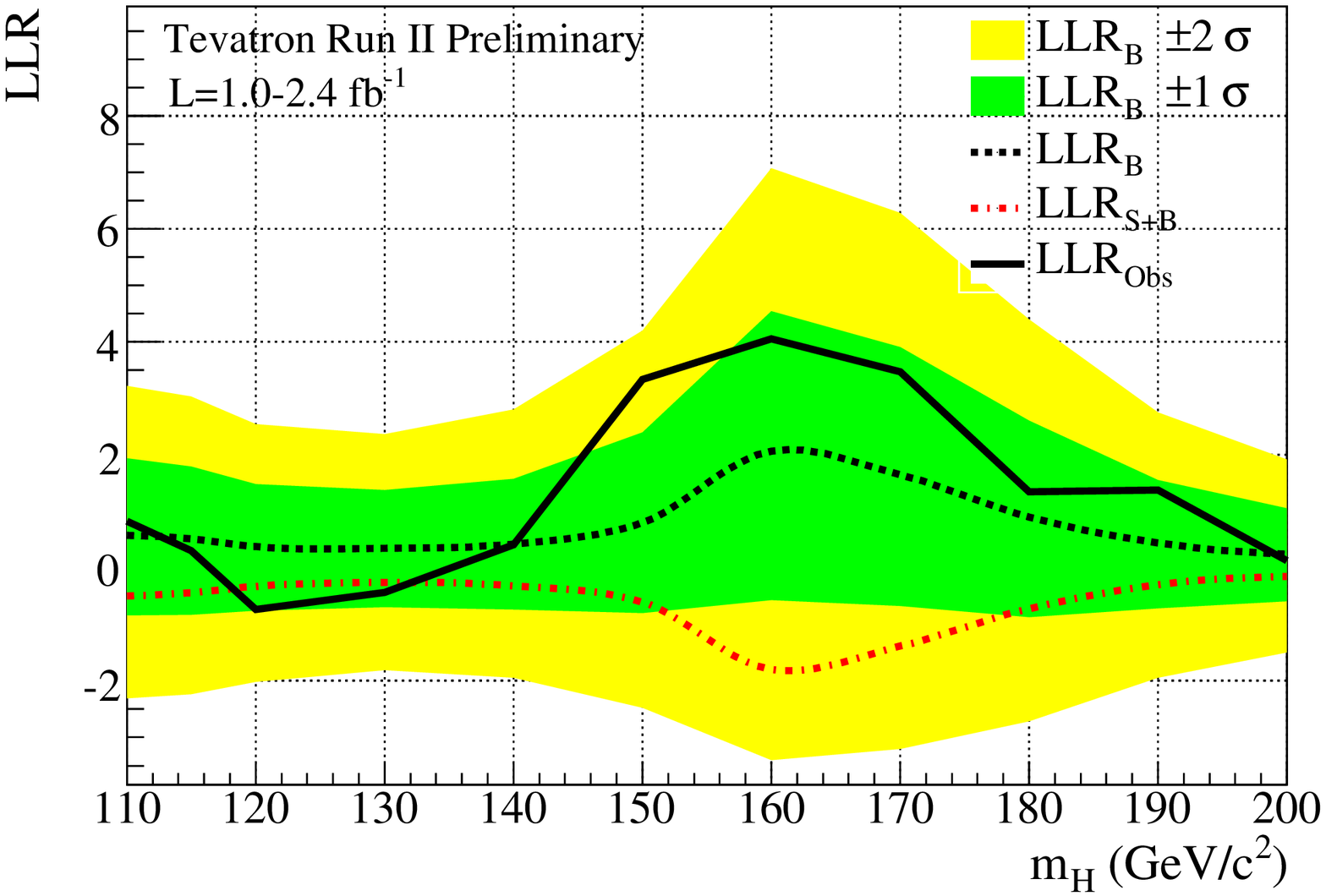}
\includegraphics[width=95mm]{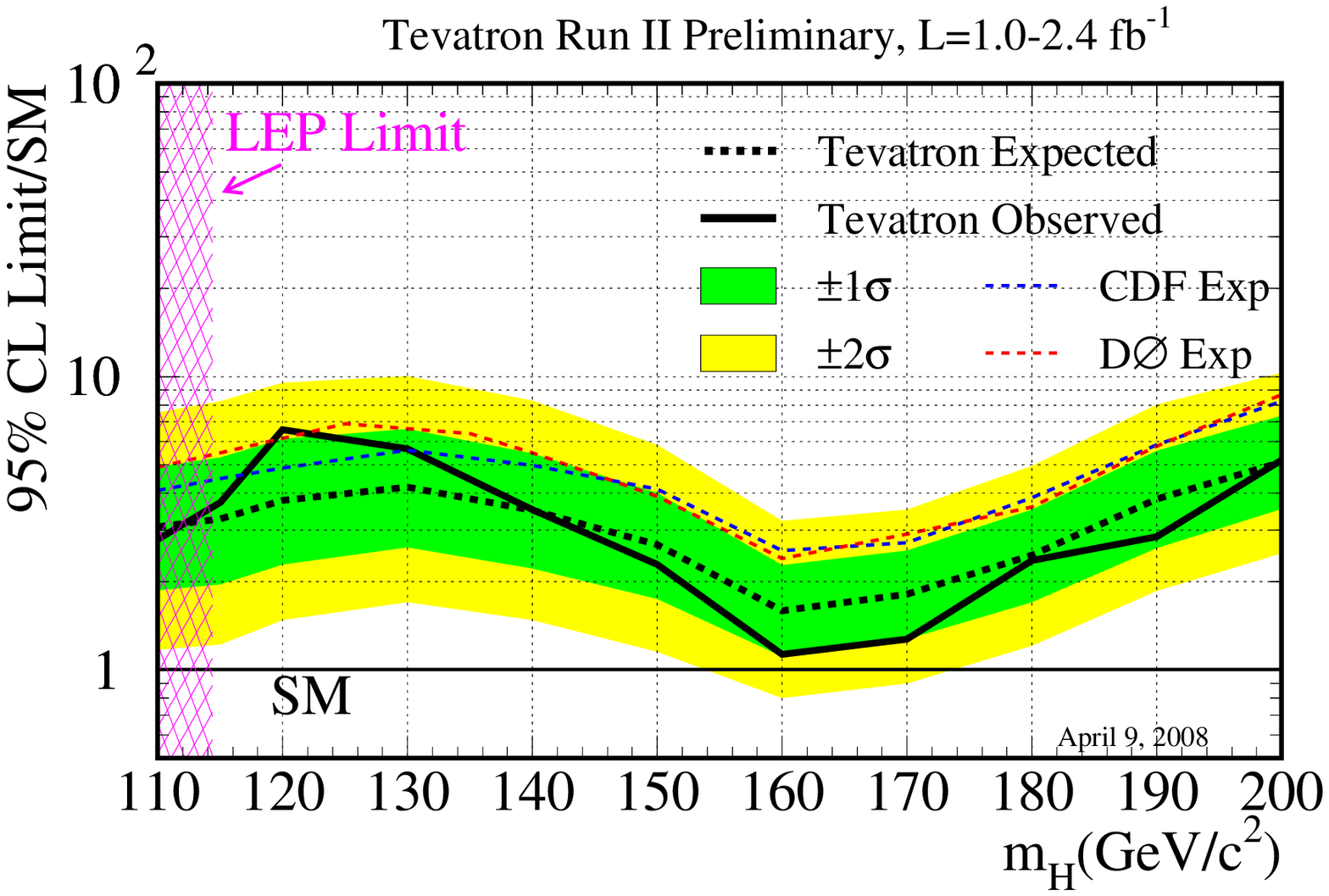}
\caption{
a) Log-Likelihood-Ration (LLR) plot
for the CDF-D\O\ combination, see \cite{bib:sys} for more details;
b) ratio to the SM Higgs production cross section
of the 95\% C.L. limit obtained with
the combined CDF-\DZ analyses;
the bands indicate the
68\% and 95\% probability regions where the limits can
fluctuate, in the absence of signal;
also shown are the  expected upper limits obtained for  
all combined CDF channels, and for  all combined D\O\ channels.
} 
\label{combos}
\begin{picture}(0,0)(0,0)
  \put (-205,205){\scriptsize {\bf (a)}}
  \put ( 40,205){\scriptsize {\bf (b)}}
\end{picture}
\end{figure*}

These results represent about a 40\% improvement in expected sensitivity over
those obtained on the combinations of results of each single experiment,
which yield  observed (expected) limits on the SM  ratios of 
5.0~(4.5) for CDF and 6.4~(5.5) for D\O\ at $m_{H}=115$~GeV, and of 
1.6~(2.6) for CDF and 2.2~(2.4) for D\O\ at $m_{H}=160$~GeV.

\begin{table}[ht]
\caption{\label{tab:ratios} Median expected and observed 95\% CL
cross section ratios for the combined CDF and D\O\ analyses as a function
the Higgs boson mass in GeV.}
\begin{tabular}{|lccccccccccc|}
\hline
$m_H$ (GeV)      & 110 & 115  & 120 & 130 & 140 & 150 & 160 & 170 & 180 & 190 & 200\\ \hline 
Expected         & 3.1 &  3.3 & 3.8 & 4.2 & 3.5 & 2.7 & 1.6 & 1.8 & 2.5 & 3.8 & 5.1\\
Observed         & 2.8 &  3.7 & 6.6 & 5.7 & 3.5 & 2.3 & 1.1 & 1.3 & 2.4 & 2.8 & 5.2\\
\hline
\end{tabular}
\end{table}

\section{PROSPECTS} 

\begin{figure*}[htbp]
\centering
\includegraphics[width=75mm]{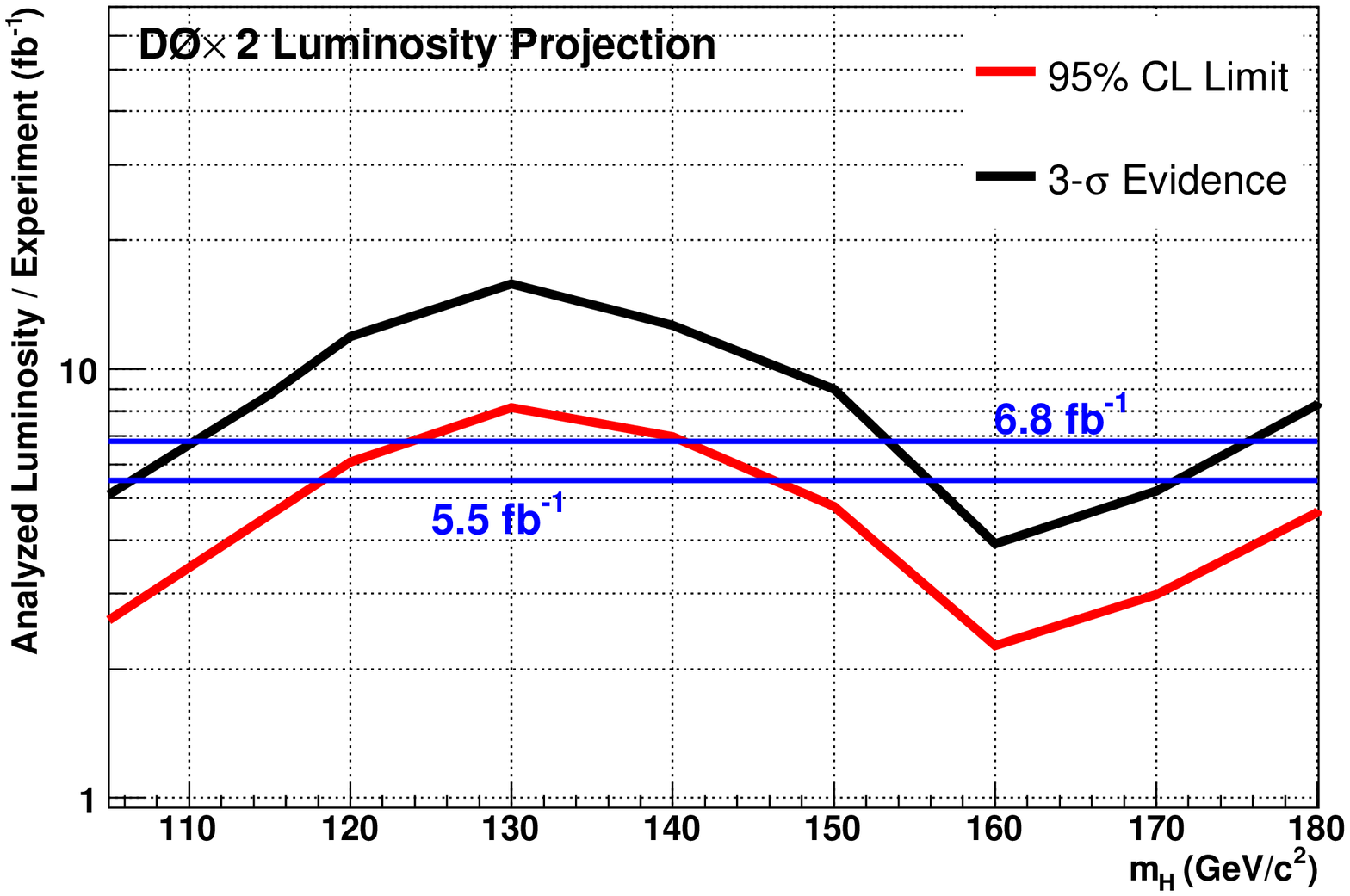}
\includegraphics[width=75mm]{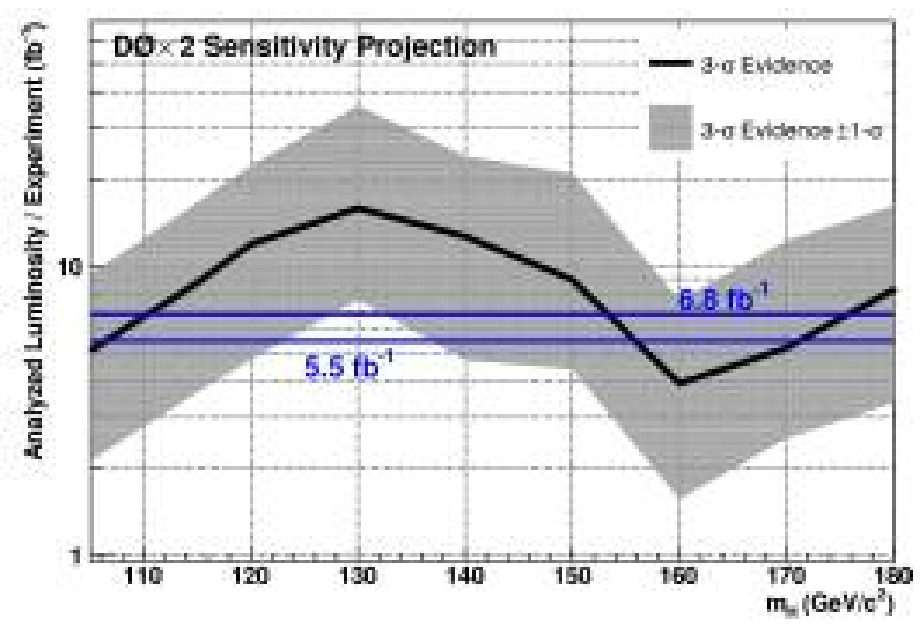}
\caption{Prospect for Higgs search sensitivity at the Tevatron 
assuming the analysis improvements described in the text and similar performance between
CDF and D\O\ analyses. The two horizontal limes correpond to a running until 2009 and 2010.} 
\label{prospects}
\begin{picture}(0,0)(0,0)
  \put (-185,165){\scriptsize {\bf (a)}}
  \put ( 40,165){\scriptsize {\bf (b)}}
\end{picture}
\end{figure*}

In Fall 2007, the two collaborations were asked to review
their prospects for Higgs sensitivity in case of a 2009 or a 
2009+2010 running of the Tevatron.
Looking back, 
since 2005, the high mass experimental sensitivity for the Higgs
has
improved by a  factor of 1.7 (without counting the
gain due to luminosity) and this was mainly due to improvements
with multivariate techniques and with lepton acceptance.
From 2007 to 2010, \DZ estimated a possible additional improvement in 
analysis sensitivity by  a factor of 1.4, coming from 
increased lepton efficiency (10\% per lepton) and
further multivariate analyses improvements (30\% in sensitivity)
Other potential improvements (like the inclusion of
tau channels) were not included in  the estimate.
CDF estimates were similar, so we could effectively estimate the
Tevatron potential independently and both
experiments reached similar conclusions.

At low mass, since 2005, the analyses sensitivities have also improved
by a factor of approximately 2,
due to increase in acceptance, kinematic phase space,
trigger efficiency, asymmetric tagging for double b-tags,
b-tagging improvements (NN b-tagging), and
improved statistical techniques/event NN discriminant
From 2007 to  2010, we estimated that we will 
gain an additional factor of 2.0 beyond the 
improvement expected from the increased luminosity:
b-tagging improvements,
di-jet mass resolution,
increased lepton efficiency,
improved multivariate techniques.

With such improvements, 
we obtain the
95\% C.L. and 3 $\sigma$ sensitivity curves given in
Figure~\ref{prospects}a as a function of the analyzed data.
The predicted delivered luminosity from the
Tevatron when adding  a 2010 running  is 8 fb$^{-1}$, resulting
in 6.8 fb$^{-1}$ of high quality analyzed data, as shown in the
Figure. 
With the data accumulated by the end of 2010,
the combined data of the CDF and \DZ experiments 
could allow to explore 
much  of the SM Higgs mass region allowed by the 
precision electroweak measurements.
Three-sigma evidence for a Higgs is possible over almost the entire range,
as can be seen from the 1-$\sigma$ band in Figure~\ref{prospects}b,
and is probable for the low and high end of this mass spectrum if the
Higgs boson lies there.
The low mass searches are particularly important, since the Tevatron
and LHC would be searching for it in different decay modes 
(\hbb\ at the Tevatron, $H \rightarrow \gamma\gamma$ at the LHC).

In conclusion, the search for the SM Higgs boson has reached a mature
state at the Tevatron, the combination techniques are in place,
and the analyses are continuing to  improve their individual
sensitivity such that the  Tevatron combination would
reach SM Higgs sensitivity between 115 and 185 GeV by the end of Run II.



\begin{acknowledgments}
The author wish to thank the organizers for a stimulating conference, and all the CDF and D\O\
colleagues for the  results obtained together and presented here.
\end{acknowledgments}

\end{document}